\documentclass[prb,twocolumn]{revtex4}
\usepackage{subfigure}
\usepackage{color}
\usepackage{array}
\usepackage{amsmath}
\usepackage{amssymb}
\usepackage{wasysym}
\usepackage{bm}
\usepackage{graphicx}
\usepackage{placeins}
\usepackage{hyperref}

\def\k{\ensuremath \bm{k}}

\def\px{\ensuremath \hat{\tau}_{x}}

\begin{document}

\title{Quantum oscillations in a bilayer with broken mirror symmetry: a minimal model for YBa$_2$Cu$_3$O$_{6 + \delta}$}
\author{Akash V. Maharaj${}^{1}$, Yi Zhang${}^{1}$, B.J. Ramshaw${}^{2}$, and S. A. Kivelson${}^{1}$}
\affiliation{${}^{1}$Department of Physics, Stanford University, Stanford, California 94305, USA.}
\affiliation{${}^{2}$Los Alamos National Laboratory, Los Alamos, New Mexico 87545, USA.}
\date{\today}

\begin{abstract}
Using an exact numerical solution and semiclassical analysis, we
investigate quantum oscillations (QOs) in a model of a bilayer
system with an anisotropic (elliptical) electron pocket in each
plane. Key features of QO experiments in the high temperature
superconducting cuprate YBCO can be reproduced by such a model, in
particular the pattern of oscillation frequencies (which reflect
``magnetic breakdown'' between the two pockets) and the polar and
azimuthal angular dependence of the oscillation amplitudes. However,
the requisite magnetic breakdown  is possible  only  under the
assumption that the horizontal mirror plane symmetry is
spontaneously broken and that the bilayer tunneling, $t_\perp$, is
substantially renormalized from its `bare' value. Under the
assumption that $t_\perp= \tilde{Z}t_\perp^{(0)}$, where
$\tilde{Z}$ is a measure of the quasiparticle weight, this suggests
that $\tilde{Z} \lesssim 1/20$. Detailed comparisons with new
YBa$_2$Cu$_3$O$_{6.58}$ QO data, taken over a very broad range of
magnetic field, confirm specific predictions made by the breakdown
scenario.
\end{abstract}

\maketitle

Quantum oscillations (QOs) are a spectacular consequence of the
presence of a Fermi surface. Their observation in the high $T_c$
cuprate
superconductors\cite{Proust2007,leboeuf2007electron,bangura2008small,yelland2008quantum,vignolle2008quantum,audouard2009multiple,sebastian2010compensated,singleton2010magnetic,ramshaw2011angle,sebastian2012quantum,sebastian2012towards,barivsic2013universal,sebastian2014normal,sebastian2015quantum,ramshaw2015quasiparticle}
combined with recent observations of charge density wave
correlations\cite{tranquada1994simultaneous,tranquada1995evidence,fujita2004stripe,howald2003periodic,hoffman2002four,hanaguri2004checkerboard,fink2009charge,laliberte2011fermi,chang2012direct,ghiringhelli2012long,RexsYBCO,doiron2013hall,leboeuf2013thermodynamic,he2014fermi,tabis2014charge,forgan2015nature},
have led to a compelling view of the  non-superconducting ``normal''
state of the underdoped cuprates at high fields, $H>H_c$, and low
temperatures, $T\ll T_c$. In this regime, small electron-like Fermi
pockets arise from reconstruction of a larger hole-like Fermi
surface presumably due to translation symmetry breaking in the form of
bidirectional \footnote{``Bidirectional'' here refers to both$(Q,0)$
and $(0,Q)$ charge density wave order parameters occurring
simultaneously in the same domain. Such a state could preserve $C_4$
symmetry, as in ``checkerboard order,'' or, as in the proposed
``criss-cross stripe'' phase  of Ref. \onlinecite{Maharaj2014},
could break this and other point-group symmetries. This is to be
contrasted with  unidirectional or stripe order where a single wave
vector CDW occurs per domain which necessarily implies breaking of
$C_4$ rotational symmetry.\cite{comin2015symmetry}}
charge-density-wave (CDW)
order\cite{Chakravarty2001,millis2007antiphase,harrison2011protected,Eun2012,harrison2012fermi,lee2014amperean,allais2014connecting,wang2014charge,Maharaj2014,russo2015random, briffa2015fermi,harrison2015nodal}.

However, to date, no theory of Fermi-surface reconstruction by a
simple CDW can simultaneously account for the Fermi pockets and the
relatively small magnitude of the measured specific
heat,\cite{yao2011fermi, riggs2011heat} which  presumably reflects
the persistence  a pseudo-gap that removes other portions  of the
original (large) Fermi surface. \footnote{More complex states in
which CDW order coexists with other orders, including an
incommensurate dDW\cite{Eun2012,wang2015onsager} and a CDW in a FL*
phase \cite{chowdhury2015higgs,chowdhury2014Density}, have been
proposed which may offer a way  to reconcile these observations. }
Thus, rather than trying to infer the {\it origin} of the Fermi
pockets, we explore a generic model of a single bilayer split pocket
to elucidate general features that can most easily {\it account for}
the salient features of the QOs.
\begin{figure}
\centering
\includegraphics[width=0.45\textwidth]{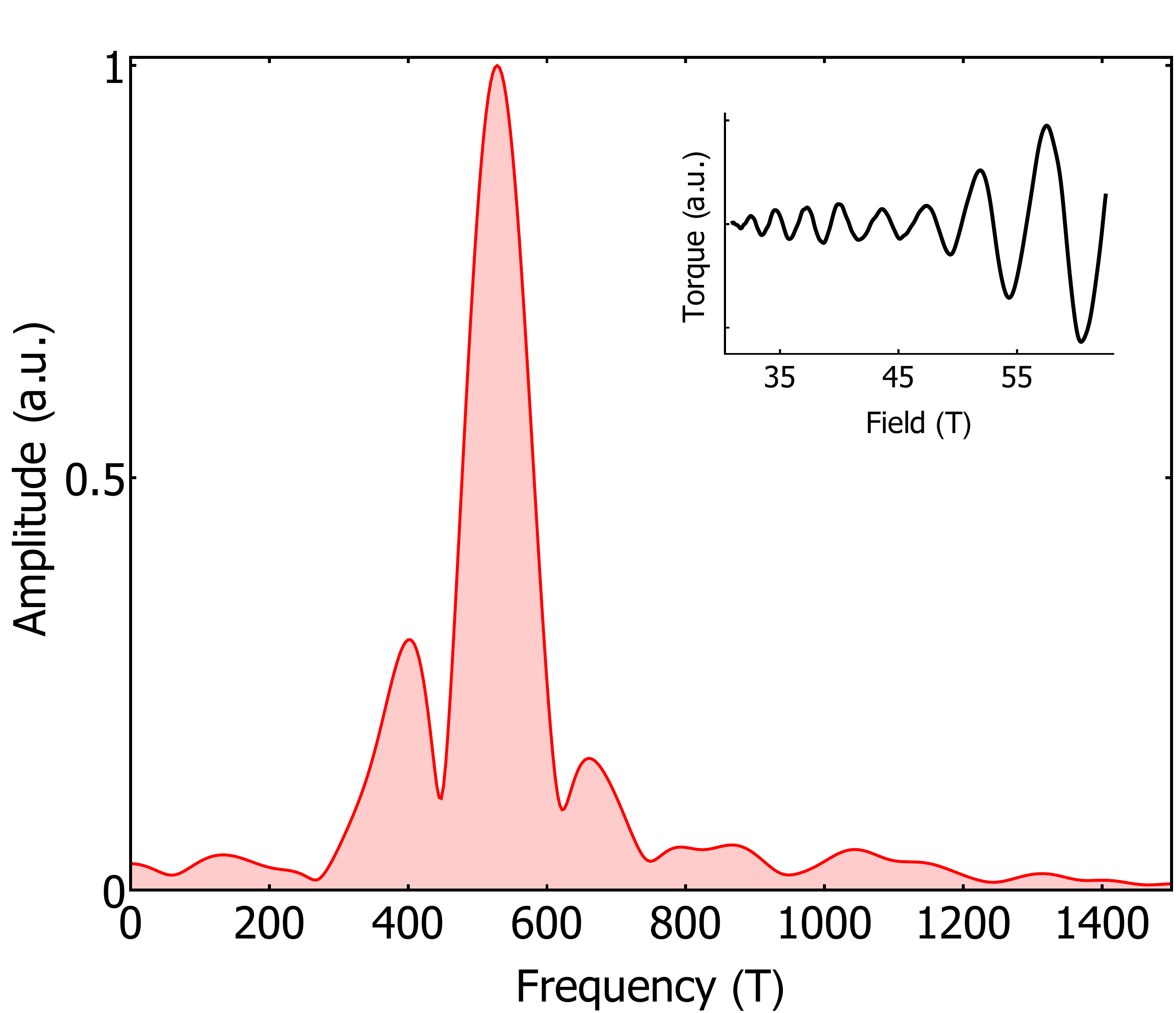}
\caption{Typical Fourier transform of QOs of the magnetic torque for
underdoped YBa${}_2$C${}_3$O$_{6.58}$ ($T_c =$ 60 K, $p\approx
11\%$) showing the characteristic symmetrically split ``three peak''
structure. Raw torque data is shown in the inset, taken at
$T\approx1.5$K, for a field range of 31 to 62.6 T. Structure above
700 T in the Fourier transform is harmonic content.}\label{exp}
\end{figure}

Specifically, we focus on the bilayer cuprate YBCO, in which quantum
oscillations have been studied in greatest detail. The frequency of the QOs and the negative values
of various relevant transport coefficients\cite{leboeuf2007electron}
establish the existence of an electron pocket enclosing an area of
order 2$\%$ of the Brillouin zone. A typical spectrum of QOs in
underdoped YBCO is shown in Fig. \ref{exp}. While there is some
suggestive evidence of more than one basic frequency---which might
suggest more than one pocket per
plane\cite{allais2014connecting,doiron2015evidence,wang2015onsager}---we
instead adopt and further elucidate a suggestion of Harrison and
Sebastian\cite{harrison2011protected,sebastian2012quantum,harrison2012fermi}
that the ``three-peak'' structure of the spectrum of oscillation
frequencies reflects magnetic breakdown orbits associated with a
single, bilayer-split pocket. In refining this suggestion, we show
that, although many aspects of the QO experiments can be
successfully accounted for in this way, the requisite magnetic
breakdown is forbidden in the presence of a mirror symmetry that
exchanges the planes of the bilayer; thus, a heretofore unnoticed
implication is that the high field phase must spontaneously break
this symmetry. Other striking features of the quantum
oscillations are the existence of prominent ``spin
zeros''\cite{ramshaw2011angle} and a strong $C_4$ symmetric
dependence of the oscillation amplitudes on the in-plane component
of the magnetic field with no evidence of the enhancement at the
``Yamaji angle'' expected from the simplest ``neck and belly''
structure of a quasi-2D Fermi
surface\cite{sebastian2014normal}.\footnote{While the Yamaji angle
is not experimentally observed close to its expected value
$\theta=59^{\circ}$, we note that if the unit cell were to be
somehow doubled in the $c$-axis direction, a Yamaji angle would be
possible near to $38^{\circ}$, which is exactly where an enhancement
is observed. Because there is no compelling evidence for $c$-axis
unit cell doubling, we ignore such a possibility in this work.
Instead, the `beat' near to $40^{\circ}$ in our numerical analysis
arises because of spin-zeroes of the satellite peaks.}

We show that all these experimental features are consistent with a
simple model in which there is an elliptical Fermi pocket in each of
the planes of a bilayer, with their principal axes rotated by
$\pi/2$ relative to each other. In terms of broken symmetries, this
is consistent with a ``criss-crossed-nematic'' \textit{component} of
whatever ordered state exists in this range of $T$ and $B$. We
assume a $\vec k$ independent coupling between the layers within a
bilayer, $t_\perp$,  and we neglect all inter-bilayer coupling,
$t_\perp^\prime\approx 0$.  As we will discuss in
Sec.~\ref{sec:discussion}, both these assumptions seem more natural
in the context of experiments and band-structure calculations of
YBCO than those made by Sebastian {\it et al.} in their pioneering
treatment of this same problem. Specifically, Sebastian {\it et al.}
assumed a strong $\vec k$ dependence associated with a presumed
vanishing of $t_\perp$ in certain crystallographic directions, a
significant role from a non-zero $t_\perp^\prime$, and broken
translation symmetry in the c-direction\footnote{We note that
recent X-ray diffraction experiments in pulsed magnetic fields up to
$30T$ have observed coherent (long ranged) charge density wave order
which \textit{does not} double the unit cell in the $c$ direction}; these do not feature in our minimal model.

Finally, we have uncovered a quantitative issue with potential
qualitative implications for magnetic breakdown. The magnitude
of $t_{\perp}$ sets the size of the gap between bilayer split Fermi
surfaces thus controlling the importance of magnetic breakdown
orbits. Because our numerical approach treats magnetic breakdown
exactly (rather than using a Zenner tunneling approach), we are
uniquely placed to examine this effect. We have found that in order
for magnetic breakdown to play a significant role in the relevant
range of $B$, it is necessary to assume that $t_\perp$ is a factor
of 20 or more smaller than its ``bare'' value $t^{(0)}_{\perp}$,
which can be estimated either from band-structure
calculations\cite{andersen1995lda,elfimov2008theory} or from angle
resolved photoemission (ARPES) studies of overdoped
YBCO.\cite{fournier2010loss}. As was emphasized both in ARPES
measurements\cite{fournier2010loss} and previous theoretical
studies\cite{chakravarty1999frustrated,ioffe1999superconductivity,carlson2000dimensional},
the ratio, $\tilde{Z}\equiv t_\perp/t_\perp^{(0)}$, is a measure of
the degree of single particle interlayer coherence, and so is
related\footnote{ While ARPES studies provide a direct measure of
the electronic spectral function, and are therefore  sensitive to
the exact quasiparticle residue $Z$, this is not the same parameter
which enters into the effective Fermi liquid parameter $t_{\perp} =
\tilde{Z}t^{(0)}_{\perp}$ in QO experiments. Here, $\tilde{Z}$ is a
measure of interlayer coherence, which in the limit of degenerate
inter-layer perturbation theory becomes exactly the quasiparticle
residue.\cite{carlson2000dimensional}} to the quasiparticle weight.
This implies that the quasiparticles responsible for the QOs are
very strongly renormalized, with $\tilde{Z} \lesssim 0.05$, which
in turn suggests that they are likely to be rather subtle, emergent
features of the high field, low temperature state. One should be cautious in interpreting higher energy or temperature
phenomena in terms of a Fermi liquid of these excitations.

\subsubsection*{Logic and Organization of the Paper}
In Sec.~\ref{sec:model}, we define an explicit lattice model of
non-interacting electrons with a band-structure engineered to
produce the desired small elliptical electron-like Fermi pockets
(shown in Fig. \ref{fig:fsone}), and describe the numerical
algorithm we have used to obtain exact results for this model as a
function of an applied magnetic field. To orient ourselves, in
Sec.~\ref{semi} we sketch the semiclassical analysis (including the
effects of magnetic breakdown) which will allow us to associate the
oscillation frequencies we will encounter with the geometry of the
Fermi surface. We then present results of the numerical analysis of
the model in Sec.~\ref{results}: In Fig. \ref{fig:rawFTs} we present
the ideal QO spectrum, while in  Fig. \ref{fig:dosevolution} we
exhibit the way in which higher harmonics are rapidly suppressed by
a non-infinite quasiparticle lifetime.  We then present spectra that
result when the range of magnetic fields analyzed is confined to
realistically accessible values, discussing both qualitative and
quantitative trends as parameters are tuned (see Fig.
\ref{fig:FTevolution}). We also study the polar and azimuthal
angular dependence of the QOs (see Fig. \ref{fig:waterfall} and
\ref{figure}), and develop accurate semiclassical arguments to
interpret our numerical results (see Figs. \ref{figure} and
\ref{figphidepend}).  Finally, in Sec.~\ref{sec:discussion} we
discuss the implications of our results for the interpretation of
experiments in the cuprates, including comparison with newly
presented QO data taken on YBa$_2$Cu$_3$O$_{6.58}$, which is used to test key features of the magnetic breakdown 
scenario discussed here. We also discuss the connection with other related theoretical work.

\section{The Model}\label{sec:model}

\begin{figure}
\begin{center}
\subfigure[ { $ t_{\perp} = 0$}]{
\includegraphics[width=0.23\textwidth]{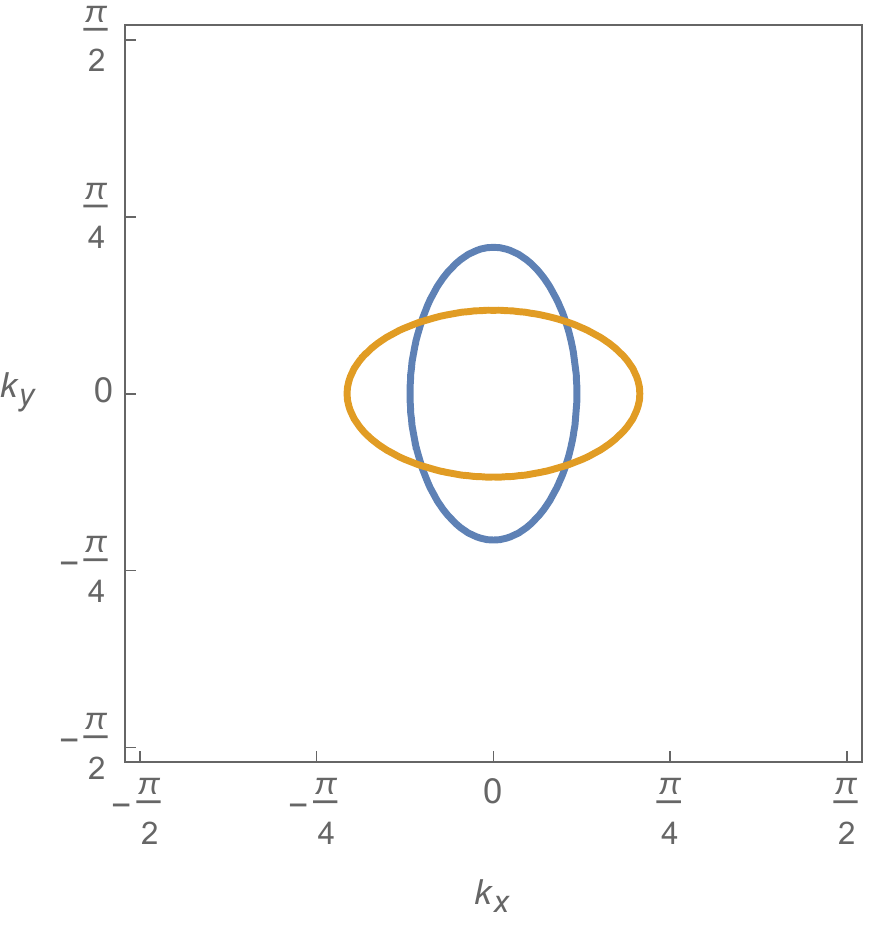}}
\subfigure[ { $t_{\perp}=0.005t_a$}]{
\includegraphics[width=0.23\textwidth]{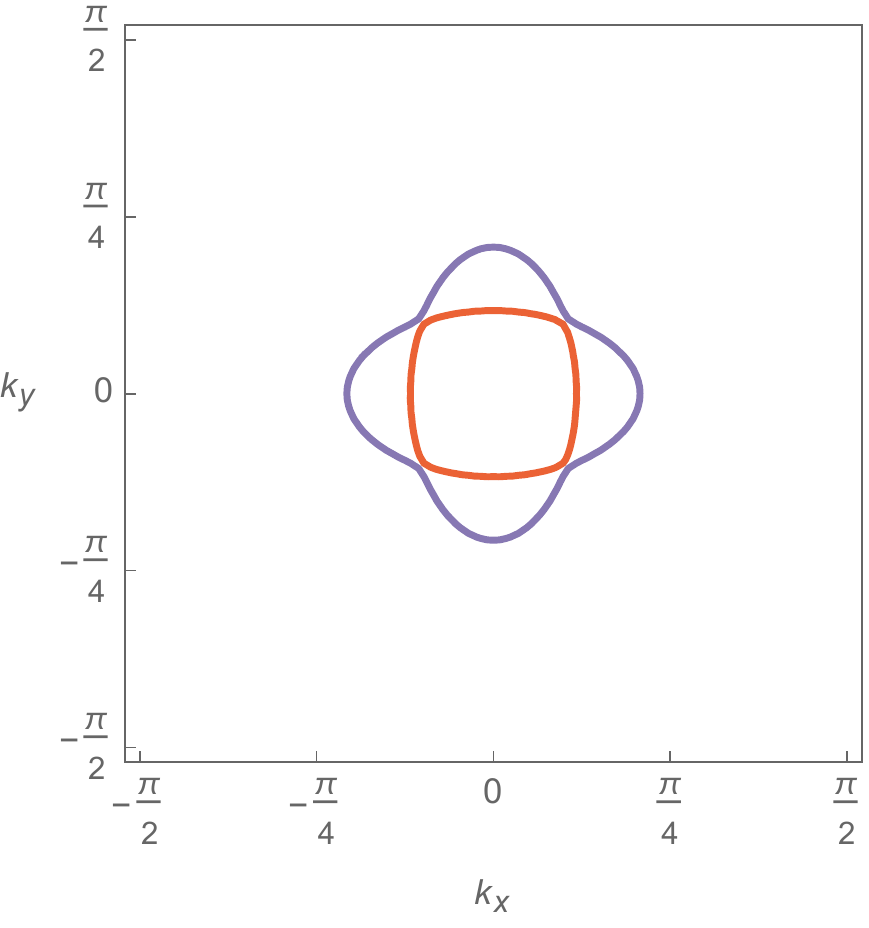}}
\caption{The Fermi surface of the bilayer system in (a) the absence
($t_{\perp} = 0$) and (b) the presence ($t_{\perp} = 0.005t_{a}$)
of an isotropic interlayer tunneling $t_{\perp}$. The
parameters used are $t_b = t_a/3$ and $\mu = -2.5306t_a$. Note that
we have zoomed in to an area that is one quarter of the full (unreconstructed)
Brillouin zone.}
\label{fig:fsone}
\end{center}
\end{figure}

We study a tight-binding model of electrons hopping on two coupled
layers, each consisting of a square lattice with purely
nearest-neighbor hopping elements. In the presence of an arbitrarily
oriented magnetic field the Hamiltonian of this model has the form
\begin{align}
\label{eq:ham}
H&= \sum_{\langle \vec{r}_{i},\vec{r}_{j} \rangle;\sigma}\sum_{\lambda} -t_{\vec{r}_i - \vec{r}_j;\lambda}\left( e^{i\Phi_{ij}} c^{\dag}_{\vec{r}_{i},\lambda,\sigma} c_{\vec{r}_{j},\lambda,\sigma} + \mbox{H.c.}\right)\nonumber\\
&+\sum_{\vec{r}_{i};\sigma} \sum_{\lambda} 4 \pi \tilde{g} B \sigma c^{\dag}_{\vec{r}_{i},\lambda,\sigma} c_{\vec{r}_{i},\lambda,\sigma} \\
&-\sum_{\vec{r}_{i};\sigma} t_{\perp}\left(
e^{i\Phi^{z}_{i}}c^{\dag}_{\vec{r}_{i},1,\sigma}
c_{\vec{r}_{i},2,\sigma} + \mbox{H.c.}\right)\nonumber
\end{align}
where $c^{\dag}_{\vec{r}_{i},\lambda,\sigma}$ is an electron creation operator at position $\vec{r}_{i}$ in layer $\lambda=1,2$ with spin $\sigma = \pm 1/2$, and  $t_{\vec{r}_i - \vec{r}_j;\lambda}$ denotes the appropriate hopping matrix element in layer $\lambda$, while $t_{\perp}$ is the (isotropic) hopping between each layer in the bilayer and $\tilde{g}$ controls the strength of Zeeman splitting. Here, $\Phi_{ij} = \int^{\vec{r}_{i}}_{\vec{r}_{j}} \bm{A}(\bm{r}) d\bm{r}$ is the phase obtained by an electron hopping from site $\vec{r}_{j}$ to $\vec{r}_{i}$
in units in which $\hbar c/e=1$, while $\Phi^{z}_{i}$ is the phase obtained upon tunneling from one layer to the next at position $\vec{r}_{i}$. To obtain perpendicularly oriented elliptical pockets we set $t_{\hat x;1} = t_{\hat y;2} = t_{a}$, and $t_{\hat y;1} = t_{\hat x;2} = t_b$. In the absence of a magnetic field this Hamiltonian can be diagonalized to give the spectrum $E_{\pm}(\k)$ where $\k = (k_x,k_y)$ is a two dimensional Bloch wavevector, with
\begin{align}
E_{\pm}(\k) &= \varepsilon_{+}(\k) \pm \sqrt{\varepsilon^{2}_{-}(\k) + t^2_{\perp}}\\
\varepsilon_{\pm}(\k) &= -(t_a \pm t_b)\cos(k_x) - (t_b \pm t_a) \cos{k_y}.
\end{align}
The Fermi surface with and without interlayer tunneling, with the choice of $t_b = t_a/3$ and a chemical potential of $\mu = -2.5306t_a$ is shown in Fig.~\ref{fig:fsone}.

In the absence of $t_{\perp}$, the addition of a magnetic field maps Eq.\ref{eq:ham} to two copies of the Hofstadter problem. Upon coupling the layers, and for fields at arbitrary polar ($\theta$) and selected azimuthal angles ($\phi$), we can always pick a gauge that preserves translation symmetry along the in-plane direction of the magnetic field, $\hat{e}$. This allows us to take the Fourier transform along $\hat{e}$, and map Eq.~\ref{eq:ham} to a modified Harper's equation.  For simplicity, we will consider the case in which the magnetic field lies in the $y-z$ plane, with the generalization to arbitrary orientation deferred to Appendix~\ref{app:technical}.  With
$\bm{B} = B\left(0,
\sin{\theta},\cos{\theta}\right)$, we can choose the gauge
\begin{align}
\bm{A} = \left(0, 2\pi \Phi x , -2\pi \Phi x \tan{\theta}\right),
\end{align}
where $\Phi=B\cos\theta$ is the
density of magnetic flux quanta
per $x-y$ lattice plaquette (in units in which the plaquette area is $1$).

Upon Fourier transforming the Hamiltonian in the $\hat{y}$ direction we
have $H = \sum_{k_y, \sigma} \hat{H}_{k_y, \sigma}$:
\begin{align}
\nonumber &\hat{H}_{k_y, \sigma} = \sum_{x, \lambda}\left\{ t_{\hat x,\lambda}\left(c^{\dag}_{(x+1,k_y);\lambda;\sigma}+ c^{\dag}_{(x-1,k_y);\lambda;\sigma} \right)\right.\\
&+\left. \left[ 2t_{\hat y,\lambda}\cos{\left(2\pi\Phi x - k_y\right)} + \frac{4\pi \tilde{g} \Phi \sigma}{\cos\theta} \right]c^{\dag}_{(x,k_y);\lambda;\sigma}\right\}c_{(x,k_y);\lambda;\sigma}\nonumber\\
&+ \sum_{x} t_{\perp}\left(e^{-i2\pi\Phi a_c \tan{\theta}}
c^{\dag}_{x,k_y;2;\sigma}c_{x,k_y;1;\sigma} +\mbox{H.c.}\right)
\label{eq:mainham0}
\end{align}
where $a_c$ is the ratio of inter-bilayer
spacing to the in-plane lattice constant.
Eq.~\ref{eq:mainham0} has three properties that make it particularly
amenable to numerical analysis: (i) the two spins $\sigma=\pm 1/2$
are decoupled and can be studied independently; 2) for
arbitrary (irrational) values of $\Phi$, the spectrum of $H$
are independent of $k_y$ in the thermodynamic limit\cite{zhang2015disruption}, allowing us to
suppress the $k_y$ summation; 3) the resulting one-dimensional
problem concerning $\hat{H}_{k_y, \sigma}$ is a block tri-diagonal
matrix, whose inverse (and by extension, the Green's function) can be
calculated recursively as described in
Appendix~\ref{appendix:technical}, allowing efficient evaluations of
its physical properties on system sizes as large as $L_x \sim
10^7$ sites along the $\hat x$ direction. In the remainder of the
paper, we will be presenting calculations of QOs in the density of
states (DOS) $\rho$ at chemical potential $\mu$, defined as
\begin{equation}
\begin{aligned}
\rho(\mu) &=  -\frac{1}{\pi L_x}\text{Tr}\left(\text{Im}[\hat{G}]\right)\\
&= -\frac{1}{\pi L_x}
\sum_{x,\lambda}\text{Im}[ G_{(x,\lambda),(x,\lambda)}(\mu)]
\end{aligned}
\end{equation}
where $G_{(x,\lambda),(x,\lambda)}(\mu)$ represents the diagonal
entry of the Green's function
\begin{align}
\hat{G}(\mu) &= \left[\left(\mu + i\delta\right)\hat{I} -
\hat{H}_{k_y, \sigma}\right]^{-1},\label{eq:greens}
\end{align}
The small imaginary term $i\delta$ gives a finite lifetime to the
electrons and broadens the Landau levels.

\subsubsection*{Choice of Parameters} %\label{sec:parameters}

For a range of values, the qualitative aspects of our results do not
depend sensitively on the values of most of the parameters that
enter the model (with the exception of the pattern of magnetic
breakdown, which we shall see is extremely sensitive to the value of
$t_\perp$).  However, to facilitate comparison with experiment, we
chose parameters so that the $k$-space area enclosed by the
elliptical Fermi pockets in the absence of interlayer tunneling is
$S_{0} \approx 530T = 1.91\% BZ$, the mean cyclotron effective mass
$m^{*} \sim 1.6 m_e$, and the electron's spin $g$ factor is $g=2$.
(See Appendix \ref{appendix:meff} for further discussion.) In the
absence of any direct experimental information concerning the
ellipticity of the Fermi pockets, we have arbitrarily adopted a
moderate anisotropy, $\sqrt{3}$ (i.e. the major axis of the ellipse
is $\sqrt{3}$ times larger than its minor axis.)

These considerations lead us to take $t_{b} = t_{a}/3$,  $\mu =
-2.5306t_a$,  and $\tilde{g}=0.87$.  Since all our calculations are
carried out at $T=0$, the overall scale of energies is unimportant,
but when referring to quantitative features of the electronic
structure of YBCO, we will take $t_a=400$meV, in which case a
characteristic inverse lifetime is $\delta = 0.005t_a \approx
(2\text{ ps})^{-1}$. We convert flux quanta per unit cell, $\Phi$,
into units of the actual magnetic field $B$, by using a unit cell
area of YBCO to be $\nu_{\text{unit cell}} =  3.82\text{\AA} \times
3.89\text{\AA}$. This means that $B$ is related to the flux per unit
cell (in units of the flux quantum) by $B =
(h/e)\times(\Phi/\nu_{\text{unit cell}}) \approx \Phi\times 27800$.
The values of the interlayer tunneling, $t_{\perp}$, and the inverse
lifetime $\delta$ are treated as unknowns; exploring the changes in
the QO spectrum which occur as they vary is one of the principle
purposes of this study.

\begin{figure}
\subfigure[{$\rho$ vs.$B$}]{
\includegraphics[width=0.23\textwidth]{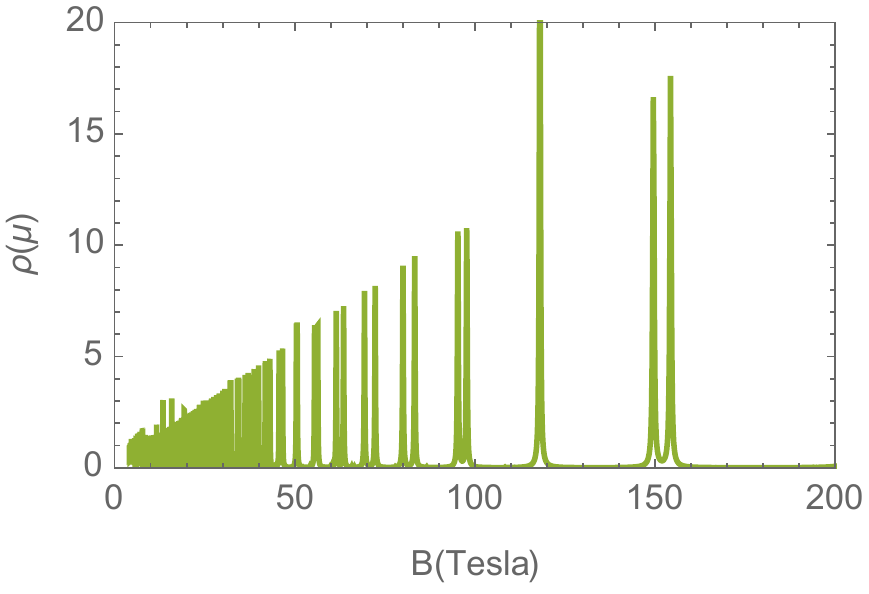}\label{fig:rawoscillationsvB}}
\subfigure[{$\rho$ vs. $1/B$}]{
\includegraphics[width=0.23\textwidth]{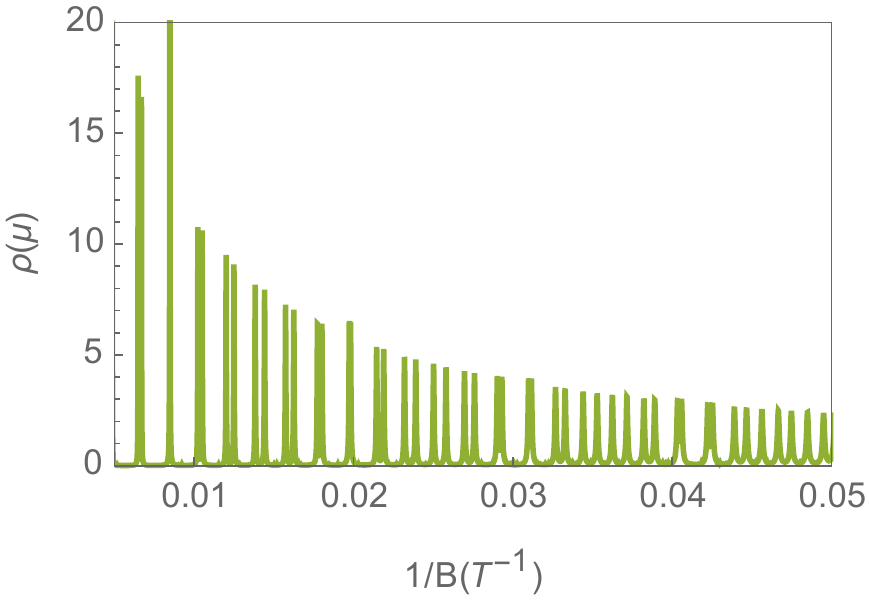}\label{fig:rawoscillationsv1ovB}} \\
\subfigure[{Fourier transform of $(b)$}]{
\includegraphics[width=0.48\textwidth]{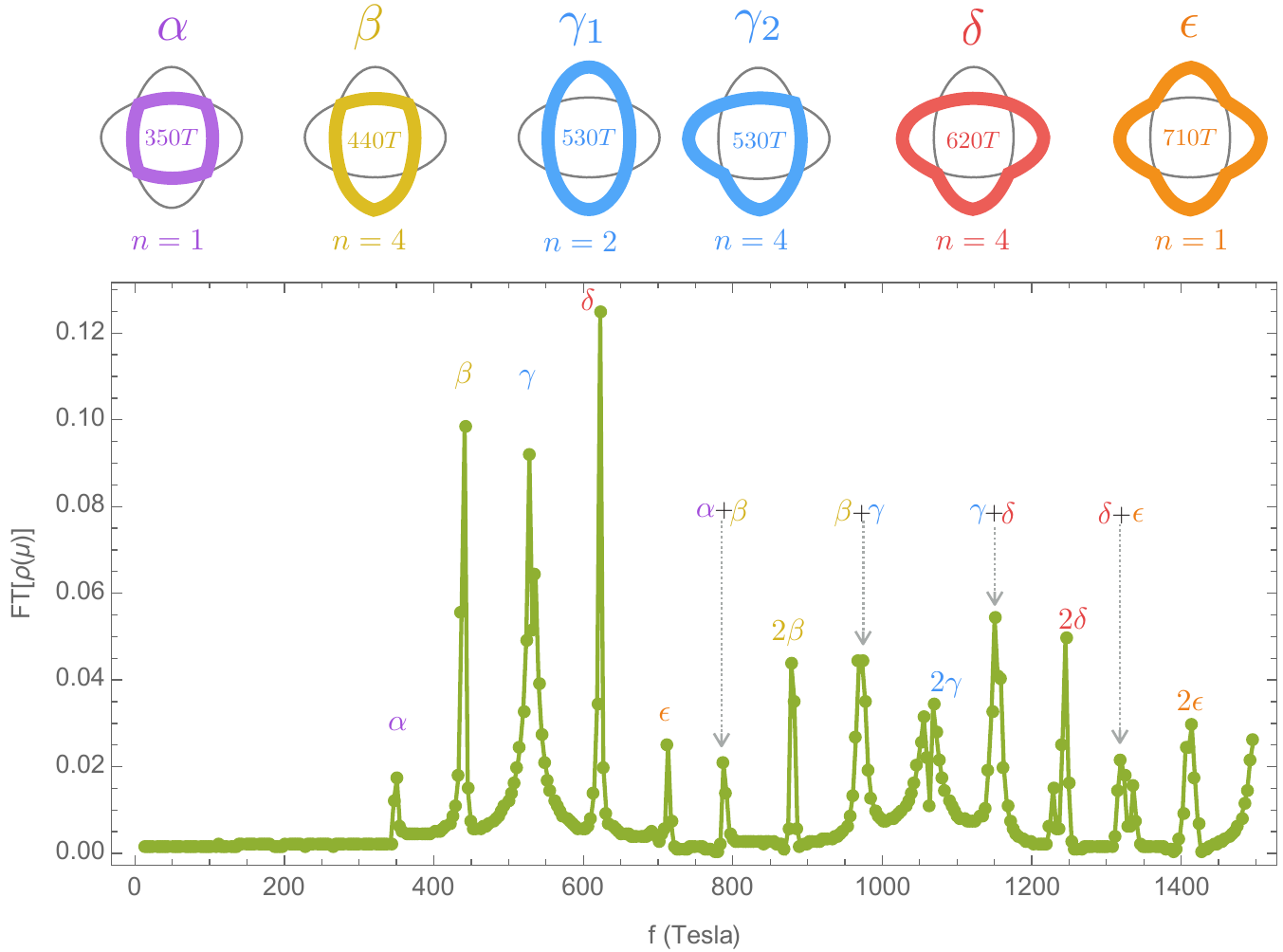}}\label{fig:rawFT}
\caption{QOs of the DOS for very small broadening $\delta=0.0001t_a$
(long lifetimes) and $t_\perp=0.005t_a$, in the absence of a Zeeman coupling ($\tilde{g}=0$).
Panels $(a)$ and $(b)$ show the calculated DOS $\rho$ vs. $B$ and
$1/B$; panel $(c)$ is the Fourier transform of panel $(b)$. Each
peak indicates a characteristic frequency of QOs and the
corresponding semiclassical orbits are also illustrated above. The number of equivalent semiclassical orbits, $n$, is indicated below each orbit, and we have 
explicitly shown the two distinct classes of $\gamma$ orbits.
 A relatively large range of magnetic field is used $4T< B <1000T$ to
capture all of the QO frequencies. The system size is $L_x =
2^{23}$. }  \label{fig:rawFTs}
\end{figure}
\begin{figure}
\begin{center}
\includegraphics[width=0.4\textwidth]{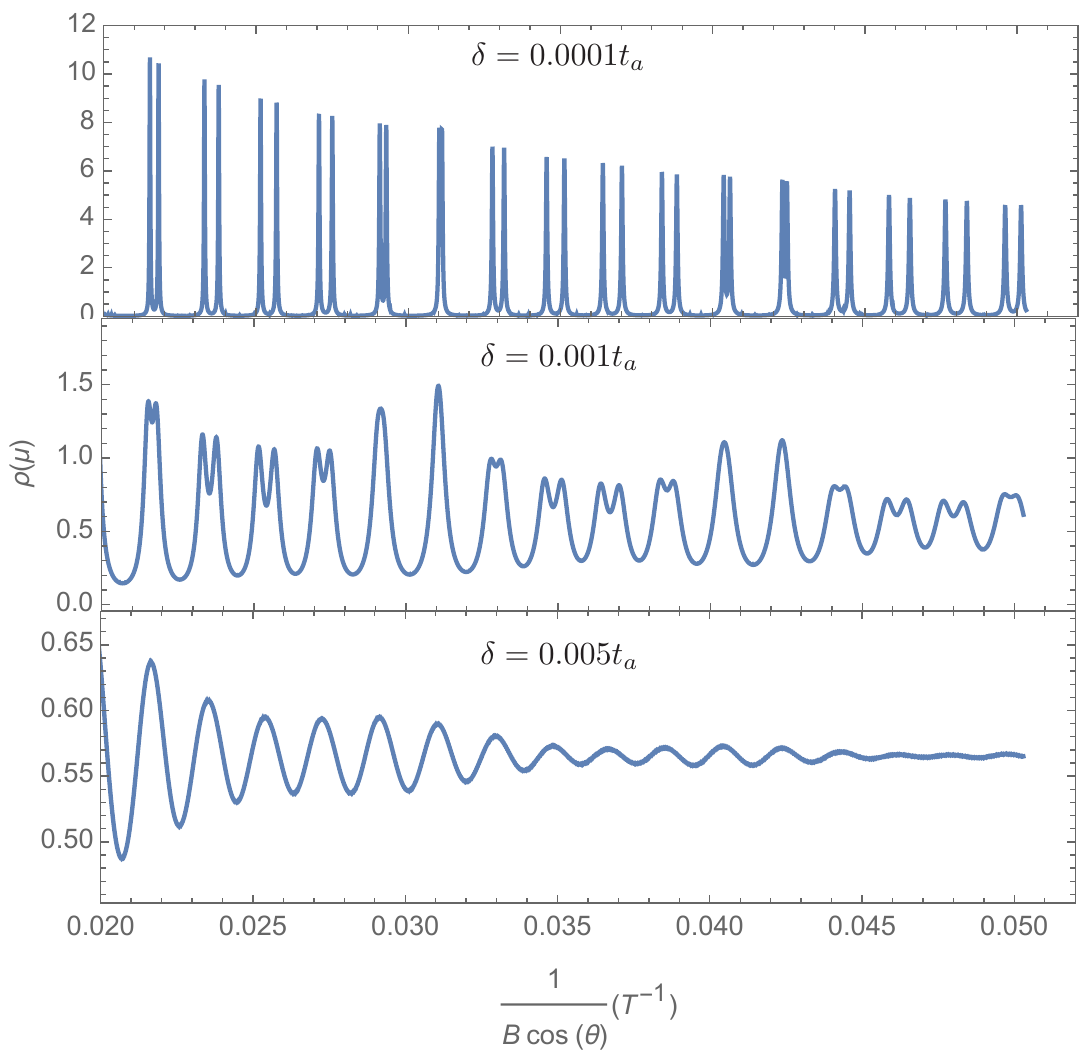}
\caption{The evolution of QOs in the DOS $\rho(\mu)$ for various values of the inverse quasiparticle lifetime, $\delta$.
The interlayer tunneling $t_{\perp} = 0.005t_a$ and the rest of the
parameters are detailed at the end of Sec.~\ref{sec:model}. }
\label{fig:dosevolution}
\end{center}
\end{figure}

\section{Semiclassical Considerations}
\label{semi}

Before undertaking the  numerical solution of this model, it is
useful to outline the results of a semiclassical analysis to
anticipate the basic structure of the QOs in the simplest situation
in which $B$ is perpendicular to the planes. As we are considering
weakly coupled bilayers, we will always assume that $t_\perp \ll
t\equiv \sqrt{t_at_b}$, so the bilayer split Fermi surfaces have
narrowly avoided crossings at four symmetry related points, as shown
in Fig. \ref{fig:fsone}b. Electrons adhere strictly to semiclassical
orbits only so long as $\hbar\omega_c \ll  t^{2}_{\perp}/t$ since
magnetic breakdown at these four points becomes significant
otherwise.  (Here $\omega_c \sim \phi t$ is the cyclotron
frequency.)  Taking this magnetic breakdown into account, there are
five distinct classes of semiclassical orbits, as shown in the
middle panel of Fig.~\ref{fig:rawFTs}, each enclosing a
$\bm{k}$-space area which, when converted  into an oscillation
frequency, correspond to five oscillation frequencies separated by
$\Delta f \approx 90T$ for the model parameters we have defined.
(These correspond to the frequencies labeled $\alpha$,
$\beta$,$\gamma$, $\delta$, and $\epsilon$ in the spectrum in the
lower panel of the figure, whose calculation is discussed in the
next section). 

The largest and smallest orbits represent the true structure of the
Fermi surface, so these two frequencies  ($\alpha$ and $\epsilon$)
must dominate the QO spectrum when $\hbar\omega_c \ll
t^{2}_{\perp}/t$. Conversely, in the limit $\hbar\omega_c \gg
t^{2}_{\perp}/t$, where to good approximation we can set
$t_\perp=0$, the spectrum is dominated by the central frequency
($\gamma_1$), in which the electron orbits are confined to  a single
plane of the bilayer, and hence correspond to the ellipses in Fig.
\ref{fig:fsone}a.\footnote{The $\gamma_2$ orbits involve tunneling from one layer 
to the next, and do not contribute in the limit of $t_{\perp} =0$.}
More complex spectra, including those with the
three peak structure seen in experiment, occur only when
$\hbar\omega \sim t^{2}_{\perp}/t$.  This, we shall see, allows us
to estimate the magnitude of $t_\perp$ directly from experiment.

We will return again to a semiclassical analysis, below, in order to
understand still more subtle features of the QO spectrum which
appear when the magnetic field is tilted relative to the Cu-O plane.

\section{Numerical results}
\label{results}
\begin{figure*}[t]
\begin{center}
\includegraphics[width=0.98\textwidth]{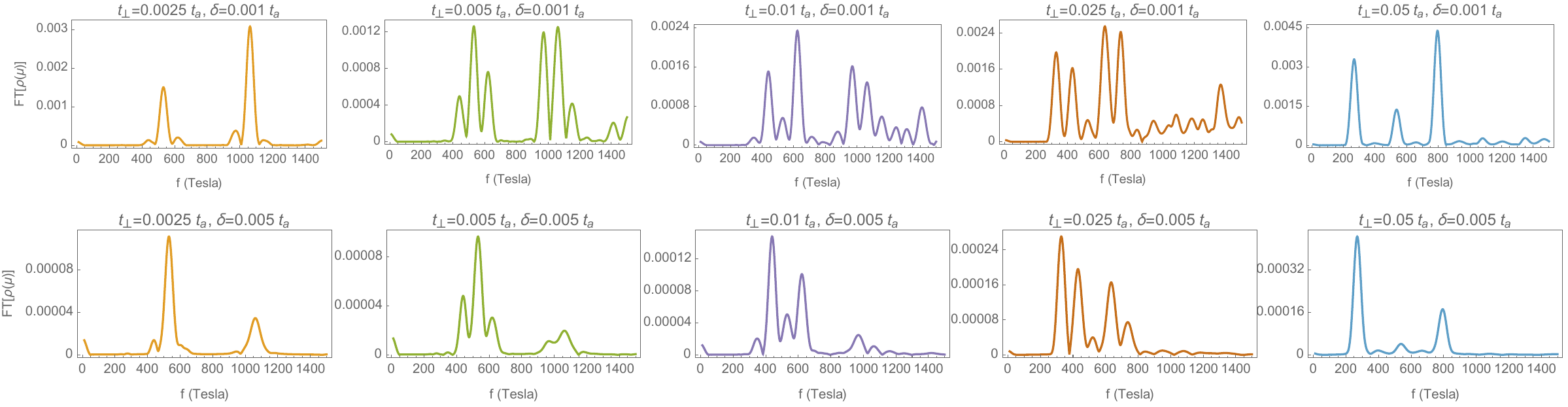}
\caption{The raw Fourier transform the density of states oscillations as the interlayer tunneling $t_{\perp}$ is increased (left to right), and the inverse lifetime $\delta$ is increased (top to bottom). Larger values of $t_{\perp}$ suppress the central frequencies and enhance the satellite frequencies which correspond to orbits of the true bilayer split Fermi surface. Shorter quasiparticle lifetimes (larger $\delta$) lead to decreased harmonic content. The field range used here is $20T < B< 100T$, with $2^{10}$ data points.}
\label{fig:FTevolution}
\end{center}
\end{figure*}

In presenting our results, we will adopt two complementary approaches. We first study an idealized theoretical limit of infinitesimally small broadening ($\delta\rightarrow 0$, i.e. infinite quasiparticle lifetime) and without  Zeeman splitting, where a sharp Landau level structure of the density of states is present and easy to interpret.
These numerical `experiments' are done for a very large range of field strengths. We subsequently study the model over an experimentally realistic range of magnetic fields with the inclusion of Zeeman splitting, while tuning the broadening and interlayer tunneling $t_{\perp}$, and subsequently examining the angular dependences. While we predominantly highlight the robust qualitative features of this model, we also focus on the quantitative aspects of magnetic breakdown, which are treated exactly in our numerical studies.

In Fig.~\ref{fig:rawFTs} we show the density of states as a function of magnetic field strengths for a $c$-directed field. The top panels show data where the broadening is infinitesimal at $\delta = 0.0001t_a$ and there is no Zeeman splitting.
Each Landau level is split due to the presence of two coupled layers, while the peaks in the density of states rise linearly with $B$ as expected for free fermions. The lower panel of Fig.~\ref{fig:rawFTs} shows the Fourier transform of this data over a large range of magnetic fields ($4T<B<1000T$). Here the high harmonic content of the oscillations is clearly seen, with comparable-in-magnitude first and second harmonics. For the first harmonics, there are five peaks clustered around a central frequency of $f = 530T$, as expected from semiclassical considerations, while at higher frequencies there are all the expected harmonic combinations giving rise to a complicated spectrum.
\subsection{Dependence on interlayer tunneling and lifetime}

We now study the model over an experimentally realistic range of
magnetic fields with a finite Zeeman coupling, $\tilde{g} =
0.87$. Fig. \ref{fig:dosevolution} and \ref{fig:FTevolution} show
the evolution of the QOs as the interlayer tunneling $t_{\perp}$ and
Landau level broadening $\delta$ are varied, where we have reduced
the range of magnetic field to $20T< B < 100T$ to conform roughly
with the range explored by current experiments in YBCO. The figures
are constructed from $2^{10}$ data points. As is clear from
Fig.~\ref{fig:dosevolution} the form of the oscillations is
radically altered as the lifetime is decreased ($\delta$ in
Eq.~\ref{eq:greens} is increased), with the sharp Landau level
structure of the spectrum becoming broadened. This leads to
oscillations with little harmonic content, while the amplitude of
the oscillatory signal is also sharply suppressed.

Fig.~\ref{fig:FTevolution} shows the Fourier transform\footnote{ In
order to smoothen the Fourier transform, we have added another
$2^{14}$ zeroes to the ends of the data, and also employed a Kaiser
window function with width parameter $\alpha= 2$ to eliminate
ringing effects. We emphasize that this procedure introduces no
additional harmonic content to the data.} of $\rho$ as both the
interlayer tunneling $t_{\perp}$ is increased (from left to right)
and the inverse lifetime $\delta$ is increased (from top to bottom).
Several qualitative features of the results are immediately
apparent. (1) As the inverse lifetime $\delta$ is increased (and the
oscillations of $\rho$ become less singular), the peaks in the
Fourier transform are also broadened while the higher frequency
peaks are preferentially suppressed in amplitude, leading to
oscillations with little harmonic content. This has a simple
semiclassical interpretation: higher frequency peaks correspond to
longer semiclassical orbits and so are suppressed in amplitude by
the decreasing quasiparticle lifetime.\cite{Bergemann:2003} (2) The competition between different
semiclassical orbits is sensitively controlled by the interlayer
tunneling $t_\perp$: as $t_\perp$ is increased, the gaps between bonding and anti-bonding Fermi surfaces increase, 
and the weight of QOs rapidly shifts from
the central frequency at $530T$ (corresponding to the 3rd orbit in
Fig.~ \ref{fig:rawFTs} which involves two magnetic breakdowns across
the true Fermi surface of the bilayer) to the side frequencies at
$(530\pm90)T$ (corresponding to the second and fourth semiclassical
orbits in Fig.~ \ref{fig:rawFTs}), and is eventually dominated by
the outermost frequencies at $(530\pm180)T$ (reflecting the `true'
bonding and anti-bonding Fermi surfaces of the bilayer).

Indeed, a particularly appealing feature of our approach is its
exact treatment of magnetic breakdown. The immediate quantitative
observation from Fig.~\ref{fig:FTevolution} is that maintaining the large (experimentally
observed) ratio of the amplitude of the central $530T$ frequency to
that of the satellite frequencies at $530\pm 90 T$ requires very
small values of the interlayer tunneling $t_{\perp} < 0.01t_a$. This
is at least an order of magnitude below the typical values of
$t_{\perp} \sim 0.1 t_a$ assigned by band structure
studies\cite{andersen1995lda,elfimov2008theory} and ARPES
studies\cite{fournier2010loss} on overdoped YBCO, but agrees
remarkably with ARPES measurements of the underdoped regime. We
discuss the consequences of this observation in Sec.
\ref{sec:discussion}.

\subsection{Polar angle ($\theta$) dependence of the QOs}

\begin{figure}
\begin{center}
\includegraphics[width=0.48\textwidth]{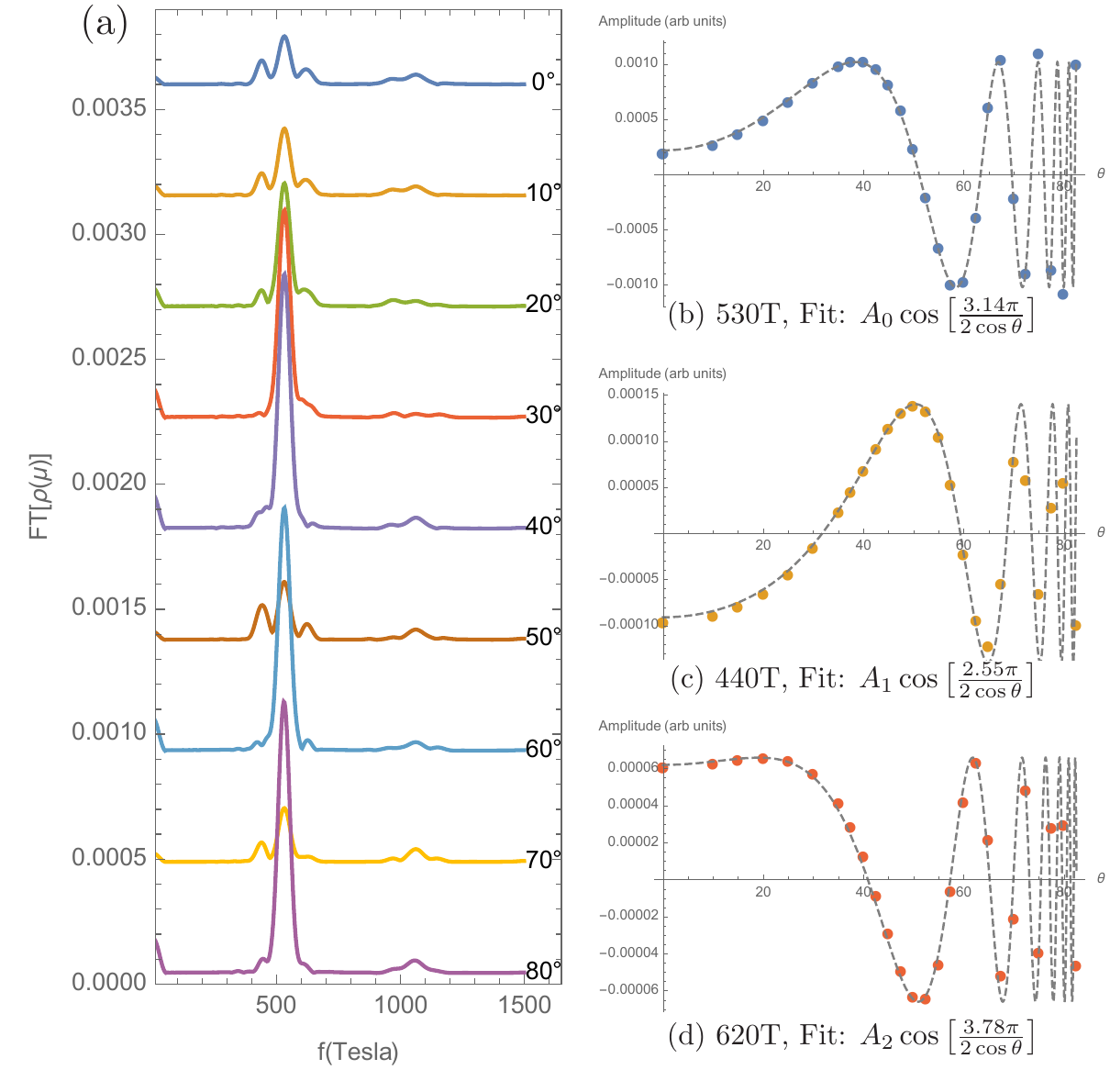}
\caption{(a) The Fourier transform of the DOS QOs for various polar angles
$\theta$ of the magnetic field $\vec B$ (different angles have been arbitrarily offset). (b)-(d)
Extracted peak heights in the Fourier transform (we have used both amplitude and phase information)
versus the polar angle $\theta$. The
parameters are $\delta = 0.005t_a$, $t_{\perp}=0.005t_a$. The dashed
lines are the theoretical fits of the angular dependence
due to spin splitting (Eq.~\ref{eq:angrho}) which is caused by Zeeman effect.}
\label{fig:waterfall}
\end{center}
\end{figure}

We now move on to cases where the magnetic field is tilted away from
the principal $c$-axis of this model and study the dependence of the
QOs on the polar and azimuthal angles,  $\theta$ and $\phi$; we also
comment briefly on the corresponding dependence seen in YBCO. Key
experimental features include the presence of spin zeroes near
$\theta \approx 51.5^{\circ}$ and $\theta \approx 63.5^{\circ}$,
with the notable absence of a Yamaji angle that is typical of simple
$s$-wave warping of a three dimensional Fermi surface. Spin zeros
(as well as the general $\theta$ dependence) arise due to Zeeman
splitting of spinful electrons. This coupling effectively shifts the
chemical potential (and hence the area of each orbit) oppositely for
each species $\sigma$, by an amount that is proportional to the
applied field $\delta f = \sigma\gamma B$. Such a $B$ dependent
shift of the Fermi surface area for each spin species becomes a
shift of the bare (spinless) frequency $f_0$ of oscillations, so
that the amplitude of oscillations for the $p$'th harmonic acquires
a field independent (but $\theta$ dependent) amplitude:
\begin{equation}
\begin{aligned}
\rho\left(\frac{1}{B},\theta\right) &\propto \sum_{\sigma = \pm 1/2} \cos{\left(2\pi p\frac{(f_0 + \sigma \gamma B)}{B\cos{\theta}}\right)} \\
&= 2\cos{\left(\frac{\pi p \gamma}{\cos{\theta}}\right)}\cos{\left(2\pi p \frac{f_0}{B}\right)}.
\label{eq:angrho}
\end{aligned}
\end{equation}
A more careful analysis shows that this field independent amplitude takes the form $A(\theta) =  \cos{\left(\pi p g\frac{m^{*}}{2m_e\cos{\theta}}\right)}$ where in practice the factor $\pi p gm^{*}/2m_e$ is related to our
definition of $\tilde{g}$ as discussed in Appendix \ref{appendix:meff}.

Fig.~\ref{fig:waterfall}(a) shows the polar angle $\theta$ dependence
of the Fourier transform of QOs for the model system in Eq. \ref{eq:ham}. The azimuthal
angle is fixed at $\phi = 45^{\circ}$ throughout the calculation. As expected,
no Yamaji-like resonance is seen because of the absence of a truly
three-dimensional dispersion.
Fig.~\ref{fig:waterfall}(b)-(d) shows the $\theta$ dependence of the QO
amplitude $A(\theta)$ at the three main frequencies. We see characteristic spin-zeroes
in the primary frequency $f=530T$ near $\theta_{0} = 51.5^{\circ}$ and
$\theta_1 = 63.5^{\circ}$. The dashed
lines show fits of the amplitude to the form given in Eq. $\ref{eq:angrho}$ -
remarkable agreement is found. We note that the positions of the spin
zeroes are different for the QOs at frequencies $440T$, $530T$ and
$620T$, despite the fact that the $g$-factor (our parameter $\tilde{g}$)
has been defined to be the same for all orbits. This robust feature of our model can be
attributed to the different effective mass of the these three orbits
which enters the form $\cos{\left(\pi p g
m^{*}/{2m_e\cos{\theta}}\right)}$, and is explored further in Appendix
\ref{appendix:meff}.

\subsection{Azimuthal angle ($\phi$) dependence of the QOs}

\begin{figure}
\begin{center}
\subfigure[{3D Cartoon plot of semiclassical orbit in the bilayer}]{
\includegraphics[width=0.45\textwidth]{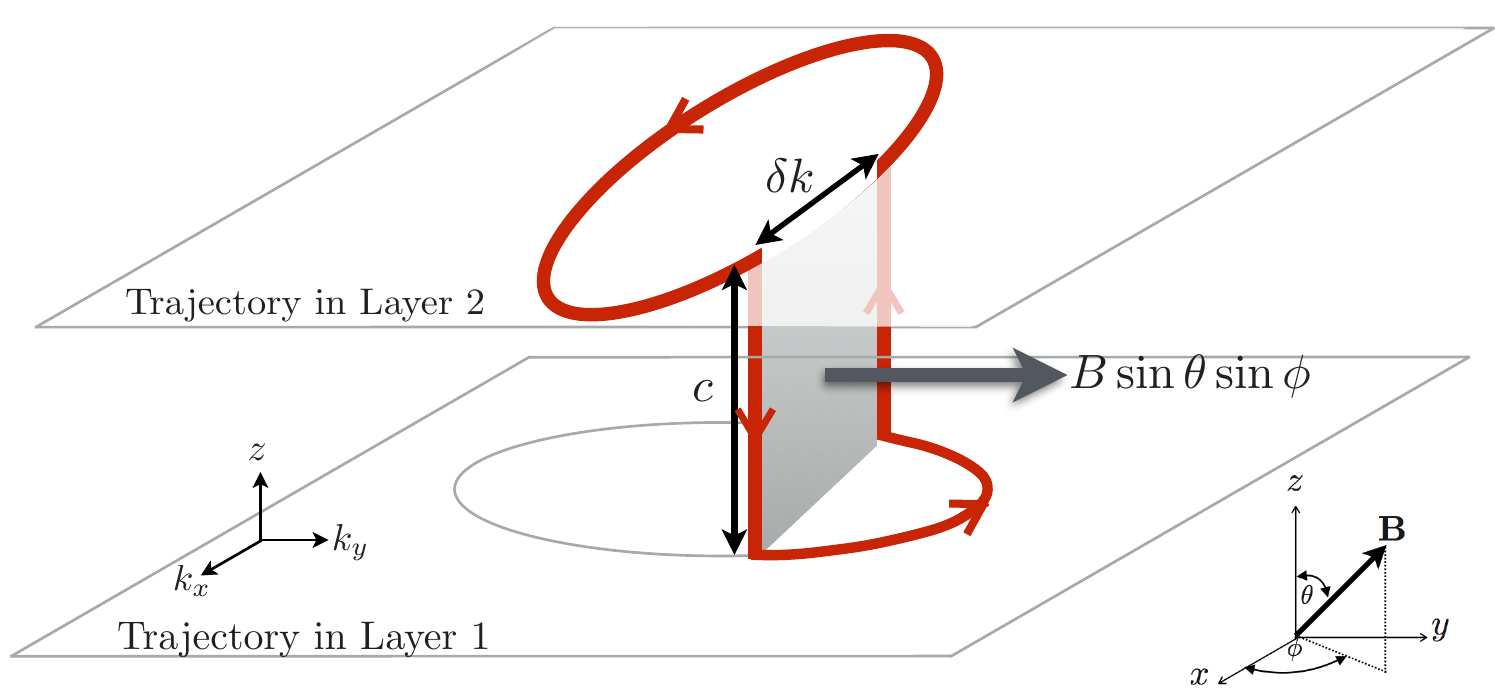}\label{fig:orbit}}
\subfigure[{Corresponding $\phi$ dependence of the QOs at $620T$ and
$440T$ given in Eq. \ref{azitheory}}]{
\includegraphics[width=0.4\textwidth]{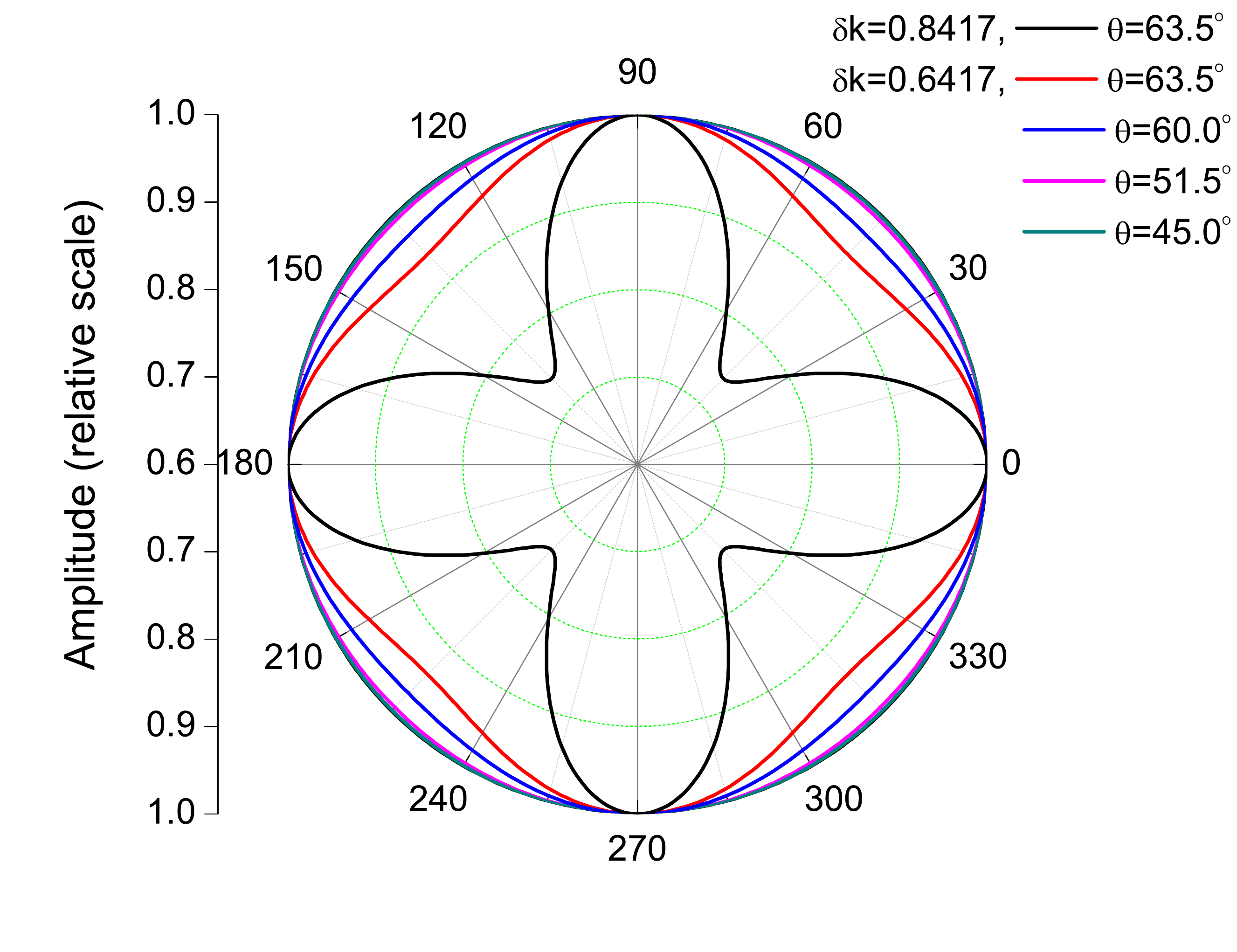}\label{figphidependtheory}}
\caption{(a) A schematic diagram showing how the in-plane component
of the flux enclosed by a given semiclassical orbit ($\delta$ orbit
in Fig. \ref{fig:rawFTs}) is determined. The red curve is the
semiclassical orbit, while grey ellipses are the Fermi surfaces.
Note that the in-plane directions are in momentum space, while the
vertical separation is in real space.  The vertical region enclosed
(shaded gray) has a (real space) area of $\delta k \ell^{2}_{B} c$.
(b) The $\phi$ dependence of QO amplitude $A_{1}(\phi)$ as in Eq.
\ref{azitheory} for various values of the polar angle $\theta$ for
$\delta k = 0.6417$ defined in our model as well as the case of a
larger $\delta k=0.8417$. The anisotropy is clearly more apparent
for larger $\theta$ and/or $\delta k$.} \label{figure}
\end{center}
\end{figure}

\begin{figure}
\begin{center}
\includegraphics[width=0.48\textwidth]{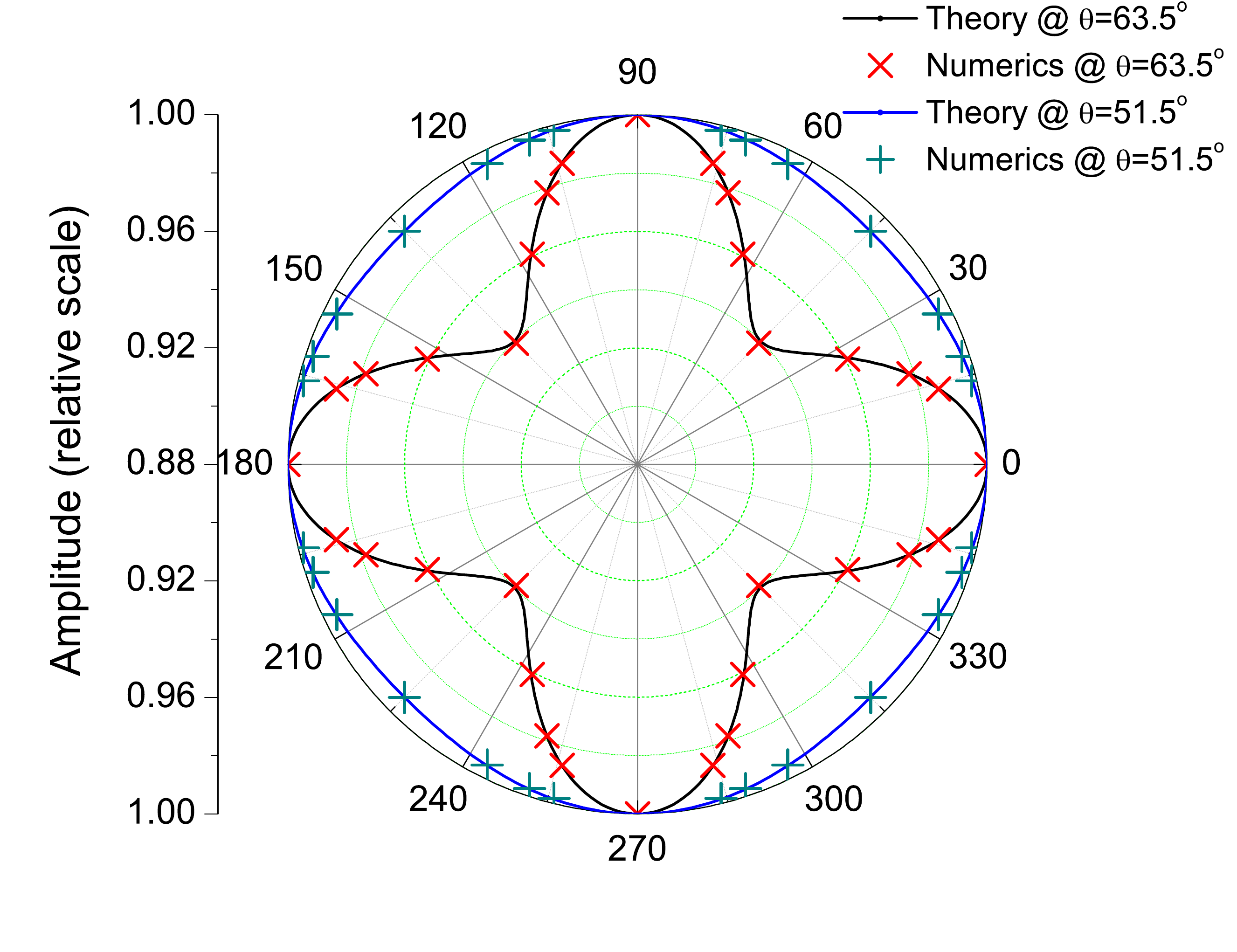}
\caption{The relative amplitude of the QOs at frequency $620T$ for
polar angles $\theta=63.5^{\circ}$ and $\theta=51.5^{\circ}$,
respectively. The QO amplitude at $\phi=0$ is maximal and set as the
unit $1$ for each of the data sets. The solid curve is the
theoretical expectation value according to Eq. \ref{azitheory} and
the expected $C_4$ rotation symmetry is clearly present. The
parameters used in our numerical calculations of the DOS QOs are
$t_{\perp}=0.005$, $\delta=0.001$ and $11T<B<100T$. }
\label{figphidepend}
\end{center}
\end{figure}

\begin{figure}
\begin{center}
\includegraphics[width=0.48\textwidth]{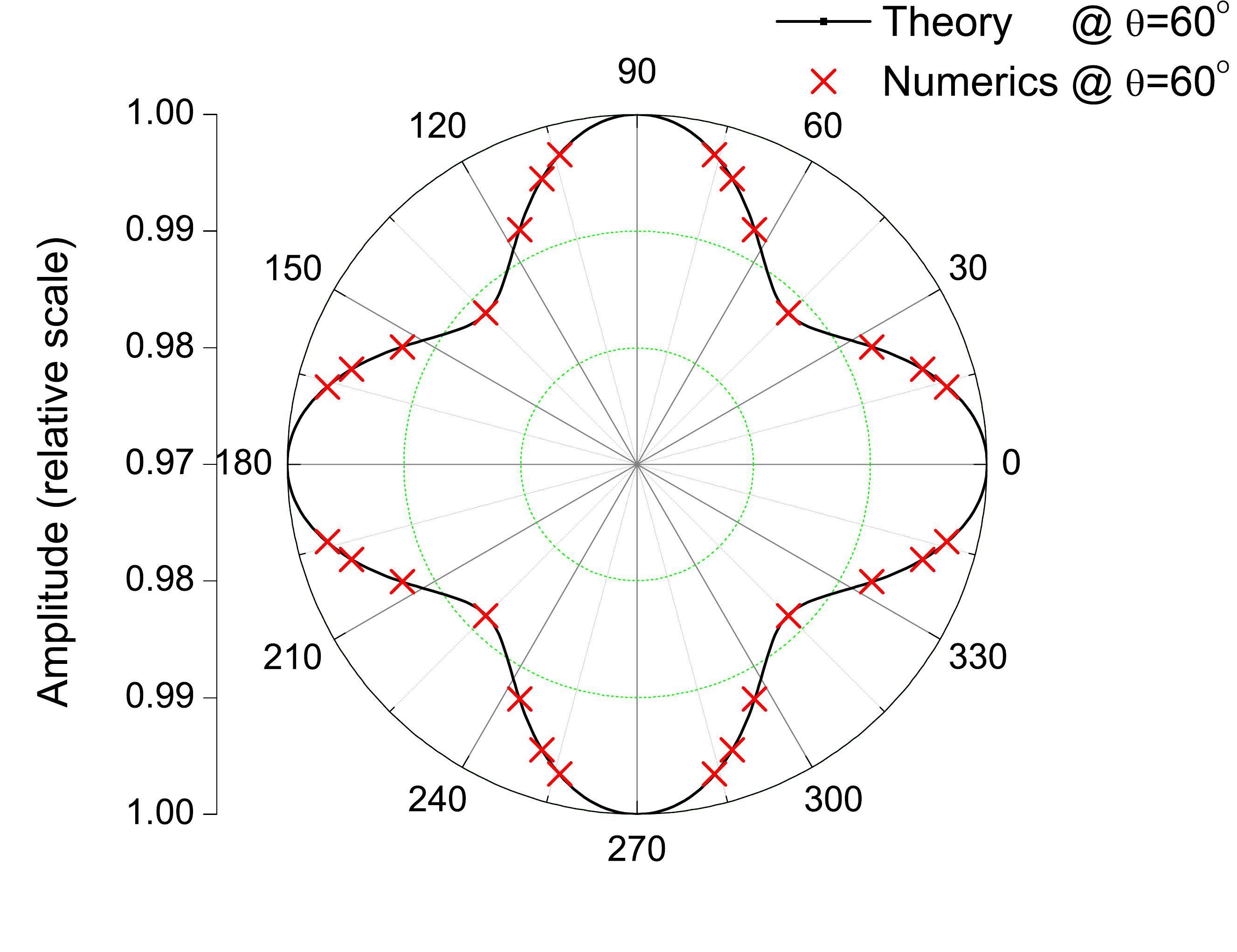}
\caption{The relative amplitude of the QOs at frequency $530T$ for
polar angles $\theta=60^{\circ}$. The solid curve is the fit to
theoretical form given by Eq. \ref{azitheory2} with $M=-13.5$. The
parameters used in our numerical calculations of the DOS QOs are
$t_{\perp}=0.005$, $\delta=0.001$ and $11T<B<100T$. }
\label{figphidepend2}
\end{center}
\end{figure}

Another notable feature of QO experiments in YBCO is the dependence
of the amplitude of the oscillations on the azimuthal angle $\phi$.
The oscillation amplitudes exhibit a four-fold anisotropy, which
increases with increasing polar angle $\theta$. Here, we show that
these features can be reproduced in our model of a single bilayer, 
with the caveat that strong anisotropy is only natural for selected orbits
that involve both layers of the bilayer ($\beta$ orbit at $440T$,
$\delta$ orbit at $620T$ and $\gamma_2$ orbit at $530T$).

In analyzing the behavior of QOs for different azimuthal angles,
much information can be gleaned from semiclassical intuitions.
First, note that the semiclassical orbits $\gamma_1$ at the central
$530T$ frequency in Fig.~\ref{fig:rawFTs} are predominantly confined
to a single layer of the bilayer. Such $2d$ orbits are only affected
by the field perpendicular to the layer, therefore no observable
azimuthal dependence is expected. On the other hand, all other
semiclassical orbits shown in Fig.~\ref{fig:rawFTs} involve
tunneling events from one layer to the next, upon which electrons
may obtain a phase proportional to the horizontal magnetic field
$B\sin{\theta}$. This means that there is weak four-fold dependence 
arising from $\gamma_2$ orbits, wherever the signal is dominated by the $530T$ frequency; conversely, a large four-fold modulation arises from the $530\pm 90T$ frequencies, and so is pronounced near to spin zero angles $\{\theta_0, \theta_1\}$ of the main $530T$ frequency.

Within the semiclassical framework, we can obtain an analytic
expression for the amplitude as a function of azimuthal angle $\phi$
by determining the amount of in-plane directed magnetic flux
enclosed by a given breakdown orbit. Fig~\ref{fig:orbit} shows the
geometry of a particular breakdown orbit for QOs at $620T$, 
where the total horizontal flux is the real space area
corresponding to the shaded region, multiplied by the field
component $B\sin{\theta}\sin{\phi}$. Semiclassically, we find the
real space area enclosed by the orbit to be  $\delta k \ell^{2}_{B}
\times c$, where $\ell^{2}_{B} = h/eB\cos{\theta}$ is the square of
the magnetic length, and $\delta k$ is the distance between the
(avoided) crossings of the Fermi surfaces (see
Fig.~\ref{fig:orbit}). Thus, the in-plane flux enclosed by this
orbit is $\Phi_{yz} = \delta k\frac{\hbar}{eB}c\times
B\tan\theta\cos\phi= c\delta k\tan\theta\sin\phi$ with our choices
of units. Similarly, there are three other possible enclosed fluxes
related by $C_4$ rotations, and given by $\Phi_{-yz} = - \Phi_{yz}$,
and $\Phi_{\pm xz} = \pm c\delta k\tan\theta\cos\phi$. The resulting
$\Phi_{j}$, $j=\pm xz, \pm yz$ each give an additional constant
initial phase to the in-plane fluxes that determine the QOs of the
corresponding reconstruction, which add up to give the overall
amplitude:
\begin{align}
A_{1}(\phi) &\propto \sum_{j} \exp{\left(i\Phi_{j}\right)}  \propto 2\cos \Phi_{xz} + 2\cos \Phi_{yz} \label{azitheory}\\
&\propto 2\cos\left(c\delta
k\tan\theta\cos\phi\right)+2\cos\left(c\delta
k\tan\theta\sin\phi\right)\nonumber
\end{align}

Examples of the azimuthal angular dependence given by Eq.
\ref{azitheory} are shown in Fig. \ref{figphidependtheory}, whose
form guarantees $C_4$ rotation symmetry. For smaller values of the
polar angle $\theta$ (and thus a smaller overall factor $c\delta
k\tan\theta$), the $\phi$ angular dependence is suppressed. We note
that the magnitude of the anisotropy depends sensitively on $\delta
k$. An example of the QO amplitude variation for a larger $\delta
k=0.8417$ as compared to the original $\delta k=0.6417$ is also
included in Fig. \ref{figphidependtheory}.

It is straightforward to verify that the angular dependence of the
QOs at $440T$ has the same result as Eq. \ref{azitheory}, while at
$530T$ we need to consider both the $\gamma_1$ and $\gamma_2$
orbits:
\begin{align}
A^\prime_{1}(\phi)  &\propto M+2\cos\left(c\delta
k\tan\theta(\cos\phi+\sin\phi)\right) \nonumber\\
&+2\cos\left(c\delta
k\tan\theta(\cos\phi-\sin\phi)\right)\label{azitheory2}
\end{align}
where $M$ is a complex constant for the contribution from $\gamma_1$
orbits and sensitively depends on the parameters including $t_\perp$
and $B$.

Another immediate consequence of this expression is that the maximum
in QO amplitudes at the side frequencies at $(530\pm90)T$ occurs
when the field is aligned with the principal axes of the ellipses.
Given that experimentally, the maximum of the oscillation amplitudes
is seen to occur for fields along the $a$ and $b$ crystallographic
directions, it is natural that the principal axes of such elliptical
pockets must lie along the $a$ and $b$ directions, i.e. such
azimuthal dependence seemingly rules out proposals where the
principal axes of the Fermi pockets are oriented at $45^{\circ}$ to
the $a$ and $b$ crystallographic directions.

Returning to the model at hand, we calculated numerically the
density of states QOs with selected values of the azimuthal angle
$\tan \phi = 0, 1, 2, 3, 4$ (plus symmetry related values) and
various polar angles. The resulting QO amplitudes at frequency
$620T$ and polar angles $\theta=63.5^{\circ}$ and
$\theta=51.5^{\circ}$ are shown in Fig. \ref{figphidepend} and is
fully consistent with the semiclassical expression derived in Eq.
\ref{azitheory}. In particular, the selected $\theta$ values are the
spin zeros of the central frequency at $f=530T$ of the QOs, where
the effect of the side frequencies at $(530\pm90)T$ are enhanced. In
addition, four-fold anisotropy is also seen for the QO amplitudes at
frequency $530T$ and polar angle $\theta=60^{\circ}$ as shown in
Fig. \ref{figphidepend2}, and fits well to Eq. \ref{azitheory2} with
parameter $M=-13.5$.

\section{
Implications for the cuprates}
\label{sec:discussion}

We have shown that a simple model of criss-crossed
elliptical electron pockets can reasonably account for the most
striking experimental observations of QOs in the bilayer cuprate
YBCO. In particular, we have shown that a three peak structure in
the Fourier transform of QOs follows naturally from the ansatz of
broken mirror symmetry\footnote{We note that such broken mirror
symmetries may be detected by probing higher rank tensorial response
functions in transport
experiments.\cite{hlobil2015elastoconductivity}} and weak bilayer
splitting. The choices of tight-binding and Zeeman-splitting
parameters that best capture this physics have been analyzed
semi-quantitatively. We have also demonstrated that major features
of both the azimuthal and polar angular dependence of the QOs can be
qualitatively reproduced by this simplified model of a single
bilayer.

A central feature of our analysis involves the small effective
interlayer tunneling $t_{\perp}$ required to account for the
prominence of the central $530T$ frequency relative to those at
$530\pm90T$.  In certain situations, a singular $k$
dependence\cite{andersen1995lda,chakravarty1993interlayer,elfimov2008theory}
of the bare interlayer tunneling, $t_\perp^{(0)}(k)\approx
t_\perp^{(0)} (\cos{k_x} - \cos{k_y})^2$, arises due to the local
quantum chemistry.  In this case  the small value of the effective
$t_\perp$  could reflect the location of the electron pockets along
the ``nodal'' direction  in the Brillouin zone where $|k_x|=|k_y|$,
rather than  any non-trivial many-body effect. However, there are
strong reasons to doubt that the bilayer tunneling in YBCO has such
strong $k$ dependence.  On theoretical grounds, LDA
studies\cite{andersen1995lda,elfimov2008theory} have found that the
tunneling between the `dimpled' planes of a YBCO bilayer remains
substantial even along the nodal direction with
$t^{(0)}_{\perp}(\k_{n}) \approx 120$meV, compared to an antinodal
value of $t^{(0)}_{\perp}(\k_{an}) \approx 150$meV. 

 This LDA prediction is supported by ARPES measurements on YBCO in the
overdoped regime\cite{fournier2010loss} where an almost isotropic
bilayer splitting of $\Delta \varepsilon_{\k_n} = 2t_{\perp}(\k_n) = 2Zt^{(0)}_{\perp}(\k_n)
\approx 130$meV in the nodal direction, compared to an antinodal
splitting $\Delta \varepsilon_{\k_{an}}  \approx 150$meV leads to a
near isotropic quasiparticle weight of $Z \approx 0.5$. This is in
sharp contrast to underdoped samples, where despite the
theoretical (LDA) prediction of a doping independent
$t^{(0)}_{\perp}$, the nodal bilayer splitting is difficult to
resolve. These experiments give an upper bound of the nodal quasiparticle weight in the underdoped regime
of $Z_{n}  < 0.065$, while an estimate based on the rescaled values
of the spectral weight yields $Z_{n} \approx 0.03$.  Such estimates agree remarkably well 
with our estimate of the effective value of $t_{\perp}$ necessary to account for the
QO's in underdoped YBCO. The constraint of the quasiparticle weight $\tilde{Z} \lesssim 0.05$, strongly suggests that the effective Fermi liquid parameter $t_{\perp} = \tilde{Z} t^{(0)}_{\perp} $ is
renormalized significantly downwards.

\subsection{Comparison with previous proposals}
There have been many
proposals\cite{Chakravarty2001,millis2007antiphase,harrison2011protected,Eun2012,harrison2012fermi,lee2014amperean,allais2014connecting,wang2014charge,Maharaj2014,russo2015random, harrison2015nodal,briffa2015fermi}
for the origin of the Fermi surface reconstruction in the cuprates.
Given recent observations of (seemingly
ubiquitous\cite{tranquada1994simultaneous,tranquada1995evidence,fujita2004stripe,howald2003periodic,hoffman2002four,hanaguri2004checkerboard,fink2009charge,laliberte2011fermi,chang2012direct,ghiringhelli2012long,RexsYBCO,doiron2013hall,leboeuf2013thermodynamic,he2014fermi,tabis2014charge,forgan2015nature})
incommensurate CDW order, a prime candidate for the Fermi surface is
one where nodally located electrons pockets are produced by
incommensurate CDWs which are at least bi-axial, involving ordering
at $\vec{Q}_x = (Q,0,1/2)$ and $\vec{Q}_y = (0,Q,1/2)$. This idea,
along with the invocation of breakdown orbits due to bilayer
splitting to account for the three peak structure of the Fourier
transform, was first advanced by Harrison, Sebastian and
co-workers\cite{harrison2011protected,sebastian2012quantum}.  In
this scenario, a diamond shaped, nodally located electron pocket is
split by bilayer tunneling (with the above mentioned $(\cos{k_x} -
\cos{k_y})^2$ form factor), with all three observed frequencies
involving orbits where the electron tunnels from one layer to the
next. The nodal location also serves to suppress simple isotropic
($s$-wave) hoping in the $c$ axis direction, leading to an absence
of a Yamaji resonance.

The model discussed in the present paper, while similar in spirit to
that of Harrison and Sebastian, possesses crucial differences of
symmetry and effective dimensionality. Under the assumption that QO
experiments probe the physics of a single bilayer, mirror symmetry
between the two layers of this bilayer must be broken in order for
breakdown orbits to be present in a purely $c$-axis directed
magnetic field -- otherwise a conserved bilayer parity associated
with the split Fermi surfaces would prevent all magnetic breakdown
(see Appendix~\ref{sec:mirrorsymmetry}). Indeed, there is evidence 
for such broken symmetry in the low field charge order.\cite{forgan2015nature, briffa2015fermi}
Once mirror symmetry is broken, a natural consequence is that the central $532T$ frequency
reflects a semiclassical orbit where electrons are confined to a
single layer of the bilayer, and if so, is naturally the most prominent
in the regime of small interlayer tunneling.\footnote{Nevertheless,
this a breakdown orbit of the `true' bonding and anti-bonding Fermi
surfaces of the bilayer} We have demonstrated that the
experimental observations can be generally accounted for in the context of a minimal model of a
single bilayer. In contrast to previous proposals, this model requires no specific $3d$-structure of the Fermi surface, and makes no specific assumptions about the nature of the order that reconstructs the Fermi surface; given that recent high field X-ray scattering experiments\cite{gerber2015three} have given evidence of an unexpected, distinct high-field character of the CDW order, we view this lack of specificity as a virtue.

\subsection{Further tests from experiments in YBa$_2$Cu$_3$O$_{6.58}$}
The magnetic breakdown scenario makes two specific predictions for QO experiments in bilayer cuprates:
\begin{enumerate}
\item Oscillations taken over a sufficiently large field range should show five spectral features distributed symmetrically about the main frequency, plus multiple higher harmonics from combination orbits.
\item The weight of the various frequency components of the quantum oscillations should be field-dependent, with orbits that require fewer breakdown events dominating at low fields.
\end{enumerate}

\begin{figure}
\centering
\includegraphics[width=0.45\textwidth]{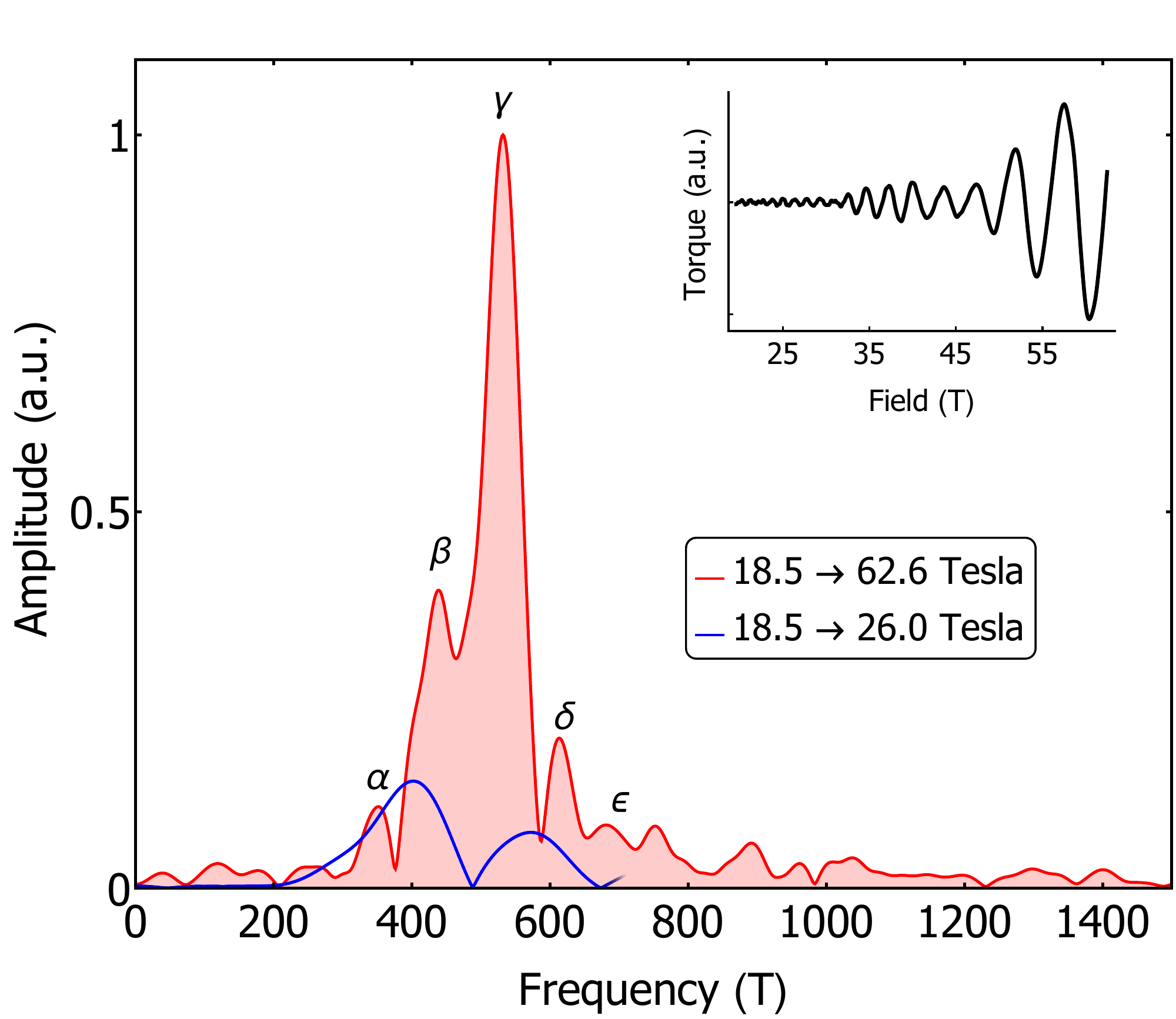}
\caption{Fourier transform of torque quantum oscillation in
YBa$_2$Cu$_3$O$_{6.58}$. Analysis of the full field range, from 18.5
to 62.6 Tesla (red curve), reveals spectral features not present in
Fig. \ref{exp}, but that correspond well with the frequencies shown
in Fig. \ref{fig:rawFTs}. Analysis of the oscillations between 18.5
and 26 Tesla only (blue curve) show that spectral weight is shifted
away from the main $\gamma$ peak, and toward the sidelobes. The blue
curve has been multiplied by a factor of 10 and truncated at 700 T
for clarity. Note that spectral features below $\approx 150$ T are
removed as part of the background-subtraction procedure, and thus
this data does not address the possibility of a 90 T frequency that
has been reported in transport measurements
\cite{doiron2015evidence}. } \label{fullrange}
\end{figure}

Fig. \ref{fullrange} shows torque magnetometry data taken on
YBa$_2$Cu$_3$O$_{6.58}$ at 1.5 kelvin. Multiple spectral components,
beyond the three main peaks identified in previous studies but
consistent with those presented in section III, are clearly visible
with this extended field range (18.5 to 62.6 Tesla). Appendix
\ref{appendix:fourier} demonstrates that these peaks (particularly
$\alpha$ and $\epsilon$) are not artifacts of the Fourier transform,
but are instead physical components of the oscillatory signal.

Transforming the data over a limited low-field range, from 18.5 to
26 T (blue curve in Fig. \ref{fullrange}), shows that the main 530 T
peak is indeed no longer dominant. Semiclassically
\cite{shoenberg:1984}, the probability of tunneling through any one
of the four junctions between the bilayer split Fermi surfaces (Fig.
\ref{fig:fsone}) is $P = e^{-B_0/B}$, where $B_0$ is the
characteristic breakdown field. The probability of avoiding
breakdown (Bragg reflection) at a junction is $(1-P)$. While this
expression is not exact (unlike the breakdown treatment in section
III), particularly at fields large compared to $B_0$, it gives
intuition as to why the spectral weight shifts at lower fields: the
$\gamma$ orbit shown in Fig. \ref{fig:rawFTs} requires four
breakdown events, while the $\alpha$ ($\epsilon$) orbit
requires none and the $\beta$ ($\delta$) orbit requires two. Note
that the field range used to obtain the blue curve in Fig.
\ref{fullrange} is insufficient to resolve the splitting of these
peaks. Finally, the dominance in amplitude of lower frequencies over
higher frequencies originates\cite{Bergemann:2003} in the suppression of larger orbits
due to quasiparticle scattering\footnote{QOs
are suppressed by a factor of $e^{-B_D/B}$ where the Dingle
reduction factor $B_D$ can be written in terms of only the
momentum-space circumference of the Fermi surface $C_F$ and the
mean-free path $l_{free}$, via $B_D = \frac{\hbar C_F}{2 e
l_free}$.}.

\section{Acknowledgements}
We acknowledge extremely useful discussions with A. Damascelli, N. Harrison, G. Lonzarich, A. P. Mackenzie, C. Proust, S. Sebastian, L. Taillefer and J. Tranquada. This work was supported in part by the US Department of Energy, Office of Basic Energy Sciences under contract DE-AC02-76SF00515 (A.V.M.), Stanford Institute for Theoretical Physics (Y.Z.), the US Department of Energy Office of Basic Energy Sciences ``Science at 100 T," (B.J.R.) and the National Science Foundation through Grant No. DMR 1265593 (S.A.K.). The National High Magnetic Field Laboratory facility is funded by the Department of Energy, the State of Florida, and the NSF under cooperative agreement DMR-1157490 (B.J.R).

\bibstyle{plain}
\bibliography{refs}

\begin{thebibliography}{62}
\expandafter\ifx\csname natexlab\endcsname\relax\def\natexlab#1{#1}\fi
\expandafter\ifx\csname bibnamefont\endcsname\relax
  \def\bibnamefont#1{#1}\fi
\expandafter\ifx\csname bibfnamefont\endcsname\relax
  \def\bibfnamefont#1{#1}\fi
\expandafter\ifx\csname citenamefont\endcsname\relax
  \def\citenamefont#1{#1}\fi
\expandafter\ifx\csname url\endcsname\relax
  \def\url#1{\texttt{#1}}\fi
\expandafter\ifx\csname urlprefix\endcsname\relax\def\urlprefix{URL }\fi
\providecommand{\bibinfo}[2]{#2}
\providecommand{\eprint}[2][]{\url{#2}}

\bibitem[{\citenamefont{Doiron-Leyraud
  et~al.}(2007)\citenamefont{Doiron-Leyraud, Proust, LeBoeuf, Levallois,
  Bonnemaison, Liang, Bonn, Hardy, and Taillefer}}]{Proust2007}
\bibinfo{author}{\bibfnamefont{N.}~\bibnamefont{Doiron-Leyraud}},
  \bibinfo{author}{\bibfnamefont{C.}~\bibnamefont{Proust}},
  \bibinfo{author}{\bibfnamefont{D.}~\bibnamefont{LeBoeuf}},
  \bibinfo{author}{\bibfnamefont{J.}~\bibnamefont{Levallois}},
  \bibinfo{author}{\bibfnamefont{J.}~\bibnamefont{Bonnemaison}},
  \bibinfo{author}{\bibfnamefont{R.}~\bibnamefont{Liang}},
  \bibinfo{author}{\bibfnamefont{D.}~\bibnamefont{Bonn}},
  \bibinfo{author}{\bibfnamefont{W.}~\bibnamefont{Hardy}}, \bibnamefont{and}
  \bibinfo{author}{\bibfnamefont{L.}~\bibnamefont{Taillefer}},
  \bibinfo{journal}{Nature} \textbf{\bibinfo{volume}{447}},
  \bibinfo{pages}{565} (\bibinfo{year}{2007}).

\bibitem[{\citenamefont{LeBoeuf et~al.}(2007)\citenamefont{LeBoeuf,
  Doiron-Leyraud, Levallois, Daou, Bonnemaison, Hussey, Balicas, Ramshaw,
  Liang, Bonn et~al.}}]{leboeuf2007electron}
\bibinfo{author}{\bibfnamefont{D.}~\bibnamefont{LeBoeuf}},
  \bibinfo{author}{\bibfnamefont{N.}~\bibnamefont{Doiron-Leyraud}},
  \bibinfo{author}{\bibfnamefont{J.}~\bibnamefont{Levallois}},
  \bibinfo{author}{\bibfnamefont{R.}~\bibnamefont{Daou}},
  \bibinfo{author}{\bibfnamefont{J.-B.} \bibnamefont{Bonnemaison}},
  \bibinfo{author}{\bibfnamefont{N.}~\bibnamefont{Hussey}},
  \bibinfo{author}{\bibfnamefont{L.}~\bibnamefont{Balicas}},
  \bibinfo{author}{\bibfnamefont{B.}~\bibnamefont{Ramshaw}},
  \bibinfo{author}{\bibfnamefont{R.}~\bibnamefont{Liang}},
  \bibinfo{author}{\bibfnamefont{D.}~\bibnamefont{Bonn}}, \bibnamefont{et~al.},
  \bibinfo{journal}{Nature} \textbf{\bibinfo{volume}{450}},
  \bibinfo{pages}{533} (\bibinfo{year}{2007}).

\bibitem[{\citenamefont{Bangura et~al.}(2008)\citenamefont{Bangura, Fletcher,
  Carrington, Levallois, Nardone, Vignolle, Heard, Doiron-Leyraud, LeBoeuf,
  Taillefer et~al.}}]{bangura2008small}
\bibinfo{author}{\bibfnamefont{A.}~\bibnamefont{Bangura}},
  \bibinfo{author}{\bibfnamefont{J.}~\bibnamefont{Fletcher}},
  \bibinfo{author}{\bibfnamefont{A.}~\bibnamefont{Carrington}},
  \bibinfo{author}{\bibfnamefont{J.}~\bibnamefont{Levallois}},
  \bibinfo{author}{\bibfnamefont{M.}~\bibnamefont{Nardone}},
  \bibinfo{author}{\bibfnamefont{B.}~\bibnamefont{Vignolle}},
  \bibinfo{author}{\bibfnamefont{P.}~\bibnamefont{Heard}},
  \bibinfo{author}{\bibfnamefont{N.}~\bibnamefont{Doiron-Leyraud}},
  \bibinfo{author}{\bibfnamefont{D.}~\bibnamefont{LeBoeuf}},
  \bibinfo{author}{\bibfnamefont{L.}~\bibnamefont{Taillefer}},
  \bibnamefont{et~al.}, \bibinfo{journal}{Physical review letters}
  \textbf{\bibinfo{volume}{100}}, \bibinfo{pages}{047004}
  (\bibinfo{year}{2008}).

\bibitem[{\citenamefont{Yelland et~al.}(2008)\citenamefont{Yelland, Singleton,
  Mielke, Harrison, Balakirev, Dabrowski, and Cooper}}]{yelland2008quantum}
\bibinfo{author}{\bibfnamefont{E.}~\bibnamefont{Yelland}},
  \bibinfo{author}{\bibfnamefont{J.}~\bibnamefont{Singleton}},
  \bibinfo{author}{\bibfnamefont{C.}~\bibnamefont{Mielke}},
  \bibinfo{author}{\bibfnamefont{N.}~\bibnamefont{Harrison}},
  \bibinfo{author}{\bibfnamefont{F.}~\bibnamefont{Balakirev}},
  \bibinfo{author}{\bibfnamefont{B.}~\bibnamefont{Dabrowski}},
  \bibnamefont{and} \bibinfo{author}{\bibfnamefont{J.}~\bibnamefont{Cooper}},
  \bibinfo{journal}{Physical Review Letters} \textbf{\bibinfo{volume}{100}},
  \bibinfo{pages}{047003} (\bibinfo{year}{2008}).

\bibitem[{\citenamefont{Vignolle et~al.}(2008)\citenamefont{Vignolle,
  Carrington, Cooper, French, Mackenzie, Jaudet, Vignolles, Proust, and
  Hussey}}]{vignolle2008quantum}
\bibinfo{author}{\bibfnamefont{B.}~\bibnamefont{Vignolle}},
  \bibinfo{author}{\bibfnamefont{A.}~\bibnamefont{Carrington}},
  \bibinfo{author}{\bibfnamefont{R.}~\bibnamefont{Cooper}},
  \bibinfo{author}{\bibfnamefont{M.}~\bibnamefont{French}},
  \bibinfo{author}{\bibfnamefont{A.}~\bibnamefont{Mackenzie}},
  \bibinfo{author}{\bibfnamefont{C.}~\bibnamefont{Jaudet}},
  \bibinfo{author}{\bibfnamefont{D.}~\bibnamefont{Vignolles}},
  \bibinfo{author}{\bibfnamefont{C.}~\bibnamefont{Proust}}, \bibnamefont{and}
  \bibinfo{author}{\bibfnamefont{N.}~\bibnamefont{Hussey}},
  \bibinfo{journal}{Nature} \textbf{\bibinfo{volume}{455}},
  \bibinfo{pages}{952} (\bibinfo{year}{2008}).

\bibitem[{\citenamefont{Audouard et~al.}(2009)\citenamefont{Audouard, Jaudet,
  Vignolles, Liang, Bonn, Hardy, Taillefer, and Proust}}]{audouard2009multiple}
\bibinfo{author}{\bibfnamefont{A.}~\bibnamefont{Audouard}},
  \bibinfo{author}{\bibfnamefont{C.}~\bibnamefont{Jaudet}},
  \bibinfo{author}{\bibfnamefont{D.}~\bibnamefont{Vignolles}},
  \bibinfo{author}{\bibfnamefont{R.}~\bibnamefont{Liang}},
  \bibinfo{author}{\bibfnamefont{D.}~\bibnamefont{Bonn}},
  \bibinfo{author}{\bibfnamefont{W.}~\bibnamefont{Hardy}},
  \bibinfo{author}{\bibfnamefont{L.}~\bibnamefont{Taillefer}},
  \bibnamefont{and} \bibinfo{author}{\bibfnamefont{C.}~\bibnamefont{Proust}},
  \bibinfo{journal}{Physical review letters} \textbf{\bibinfo{volume}{103}},
  \bibinfo{pages}{157003} (\bibinfo{year}{2009}).

\bibitem[{\citenamefont{Sebastian et~al.}(2010)\citenamefont{Sebastian,
  Harrison, Goddard, Altarawneh, Mielke, Liang, Bonn, Hardy, Andersen, and
  Lonzarich}}]{sebastian2010compensated}
\bibinfo{author}{\bibfnamefont{S.~E.} \bibnamefont{Sebastian}},
  \bibinfo{author}{\bibfnamefont{N.}~\bibnamefont{Harrison}},
  \bibinfo{author}{\bibfnamefont{P.}~\bibnamefont{Goddard}},
  \bibinfo{author}{\bibfnamefont{M.}~\bibnamefont{Altarawneh}},
  \bibinfo{author}{\bibfnamefont{C.}~\bibnamefont{Mielke}},
  \bibinfo{author}{\bibfnamefont{R.}~\bibnamefont{Liang}},
  \bibinfo{author}{\bibfnamefont{D.}~\bibnamefont{Bonn}},
  \bibinfo{author}{\bibfnamefont{W.}~\bibnamefont{Hardy}},
  \bibinfo{author}{\bibfnamefont{O.}~\bibnamefont{Andersen}}, \bibnamefont{and}
  \bibinfo{author}{\bibfnamefont{G.}~\bibnamefont{Lonzarich}},
  \bibinfo{journal}{Physical Review B} \textbf{\bibinfo{volume}{81}},
  \bibinfo{pages}{214524} (\bibinfo{year}{2010}).

\bibitem[{\citenamefont{Singleton et~al.}(2010)\citenamefont{Singleton,
  de~La~Cruz, McDonald, Li, Altarawneh, Goddard, Franke, Rickel, Mielke, Yao
  et~al.}}]{singleton2010magnetic}
\bibinfo{author}{\bibfnamefont{J.}~\bibnamefont{Singleton}},
  \bibinfo{author}{\bibfnamefont{C.}~\bibnamefont{de~La~Cruz}},
  \bibinfo{author}{\bibfnamefont{R.}~\bibnamefont{McDonald}},
  \bibinfo{author}{\bibfnamefont{S.}~\bibnamefont{Li}},
  \bibinfo{author}{\bibfnamefont{M.}~\bibnamefont{Altarawneh}},
  \bibinfo{author}{\bibfnamefont{P.}~\bibnamefont{Goddard}},
  \bibinfo{author}{\bibfnamefont{I.}~\bibnamefont{Franke}},
  \bibinfo{author}{\bibfnamefont{D.}~\bibnamefont{Rickel}},
  \bibinfo{author}{\bibfnamefont{C.}~\bibnamefont{Mielke}},
  \bibinfo{author}{\bibfnamefont{X.}~\bibnamefont{Yao}}, \bibnamefont{et~al.},
  \bibinfo{journal}{Physical review letters} \textbf{\bibinfo{volume}{104}},
  \bibinfo{pages}{086403} (\bibinfo{year}{2010}).

\bibitem[{\citenamefont{Ramshaw et~al.}(2011)\citenamefont{Ramshaw, Vignolle,
  Day, Liang, Hardy, Proust, and Bonn}}]{ramshaw2011angle}
\bibinfo{author}{\bibfnamefont{B.}~\bibnamefont{Ramshaw}},
  \bibinfo{author}{\bibfnamefont{B.}~\bibnamefont{Vignolle}},
  \bibinfo{author}{\bibfnamefont{J.}~\bibnamefont{Day}},
  \bibinfo{author}{\bibfnamefont{R.}~\bibnamefont{Liang}},
  \bibinfo{author}{\bibfnamefont{W.}~\bibnamefont{Hardy}},
  \bibinfo{author}{\bibfnamefont{C.}~\bibnamefont{Proust}}, \bibnamefont{and}
  \bibinfo{author}{\bibfnamefont{D.}~\bibnamefont{Bonn}},
  \bibinfo{journal}{Nature Physics} \textbf{\bibinfo{volume}{7}},
  \bibinfo{pages}{234} (\bibinfo{year}{2011}).

\bibitem[{\citenamefont{Sebastian
  et~al.}(2012{\natexlab{a}})\citenamefont{Sebastian, Harrison, Liang, Bonn,
  Hardy, Mielke, and Lonzarich}}]{sebastian2012quantum}
\bibinfo{author}{\bibfnamefont{S.~E.} \bibnamefont{Sebastian}},
  \bibinfo{author}{\bibfnamefont{N.}~\bibnamefont{Harrison}},
  \bibinfo{author}{\bibfnamefont{R.}~\bibnamefont{Liang}},
  \bibinfo{author}{\bibfnamefont{D.}~\bibnamefont{Bonn}},
  \bibinfo{author}{\bibfnamefont{W.}~\bibnamefont{Hardy}},
  \bibinfo{author}{\bibfnamefont{C.}~\bibnamefont{Mielke}}, \bibnamefont{and}
  \bibinfo{author}{\bibfnamefont{G.}~\bibnamefont{Lonzarich}},
  \bibinfo{journal}{Physical review letters} \textbf{\bibinfo{volume}{108}},
  \bibinfo{pages}{196403} (\bibinfo{year}{2012}{\natexlab{a}}).

\bibitem[{\citenamefont{Sebastian
  et~al.}(2012{\natexlab{b}})\citenamefont{Sebastian, Harrison, and
  Lonzarich}}]{sebastian2012towards}
\bibinfo{author}{\bibfnamefont{S.~E.} \bibnamefont{Sebastian}},
  \bibinfo{author}{\bibfnamefont{N.}~\bibnamefont{Harrison}}, \bibnamefont{and}
  \bibinfo{author}{\bibfnamefont{G.~G.} \bibnamefont{Lonzarich}},
  \bibinfo{journal}{Reports on Progress in Physics}
  \textbf{\bibinfo{volume}{75}}, \bibinfo{pages}{102501}
  (\bibinfo{year}{2012}{\natexlab{b}}).

\bibitem[{\citenamefont{Bari{\v{s}}i{\'c}
  et~al.}(2013)\citenamefont{Bari{\v{s}}i{\'c}, Badoux, Chan, Dorow, Tabis,
  Vignolle, Yu, B{\'e}ard, Zhao, Proust et~al.}}]{barivsic2013universal}
\bibinfo{author}{\bibfnamefont{N.}~\bibnamefont{Bari{\v{s}}i{\'c}}},
  \bibinfo{author}{\bibfnamefont{S.}~\bibnamefont{Badoux}},
  \bibinfo{author}{\bibfnamefont{M.~K.} \bibnamefont{Chan}},
  \bibinfo{author}{\bibfnamefont{C.}~\bibnamefont{Dorow}},
  \bibinfo{author}{\bibfnamefont{W.}~\bibnamefont{Tabis}},
  \bibinfo{author}{\bibfnamefont{B.}~\bibnamefont{Vignolle}},
  \bibinfo{author}{\bibfnamefont{G.}~\bibnamefont{Yu}},
  \bibinfo{author}{\bibfnamefont{J.}~\bibnamefont{B{\'e}ard}},
  \bibinfo{author}{\bibfnamefont{X.}~\bibnamefont{Zhao}},
  \bibinfo{author}{\bibfnamefont{C.}~\bibnamefont{Proust}},
  \bibnamefont{et~al.}, \bibinfo{journal}{Nature Physics}
  \textbf{\bibinfo{volume}{9}}, \bibinfo{pages}{761} (\bibinfo{year}{2013}).

\bibitem[{\citenamefont{Sebastian et~al.}(2014)\citenamefont{Sebastian,
  Harrison, Balakirev, Altarawneh, Goddard, Liang, Bonn, Hardy, and
  Lonzarich}}]{sebastian2014normal}
\bibinfo{author}{\bibfnamefont{S.~E.} \bibnamefont{Sebastian}},
  \bibinfo{author}{\bibfnamefont{N.}~\bibnamefont{Harrison}},
  \bibinfo{author}{\bibfnamefont{F.}~\bibnamefont{Balakirev}},
  \bibinfo{author}{\bibfnamefont{M.}~\bibnamefont{Altarawneh}},
  \bibinfo{author}{\bibfnamefont{P.}~\bibnamefont{Goddard}},
  \bibinfo{author}{\bibfnamefont{R.}~\bibnamefont{Liang}},
  \bibinfo{author}{\bibfnamefont{D.}~\bibnamefont{Bonn}},
  \bibinfo{author}{\bibfnamefont{W.}~\bibnamefont{Hardy}}, \bibnamefont{and}
  \bibinfo{author}{\bibfnamefont{G.}~\bibnamefont{Lonzarich}},
  \bibinfo{journal}{Nature} \textbf{\bibinfo{volume}{511}}, \bibinfo{pages}{61}
  (\bibinfo{year}{2014}).

\bibitem[{\citenamefont{Sebastian and Proust}(2015)}]{sebastian2015quantum}
\bibinfo{author}{\bibfnamefont{S.~E.} \bibnamefont{Sebastian}}
  \bibnamefont{and} \bibinfo{author}{\bibfnamefont{C.}~\bibnamefont{Proust}},
  \bibinfo{journal}{Annu. Rev. Condens. Matter Phys.}
  \textbf{\bibinfo{volume}{6}}, \bibinfo{pages}{411} (\bibinfo{year}{2015}).

\bibitem[{\citenamefont{Ramshaw et~al.}(2015)\citenamefont{Ramshaw, Sebastian,
  McDonald, Day, Tan, Zhu, Betts, Liang, Bonn, Hardy
  et~al.}}]{ramshaw2015quasiparticle}
\bibinfo{author}{\bibfnamefont{B.}~\bibnamefont{Ramshaw}},
  \bibinfo{author}{\bibfnamefont{S.}~\bibnamefont{Sebastian}},
  \bibinfo{author}{\bibfnamefont{R.}~\bibnamefont{McDonald}},
  \bibinfo{author}{\bibfnamefont{J.}~\bibnamefont{Day}},
  \bibinfo{author}{\bibfnamefont{B.}~\bibnamefont{Tan}},
  \bibinfo{author}{\bibfnamefont{Z.}~\bibnamefont{Zhu}},
  \bibinfo{author}{\bibfnamefont{J.}~\bibnamefont{Betts}},
  \bibinfo{author}{\bibfnamefont{R.}~\bibnamefont{Liang}},
  \bibinfo{author}{\bibfnamefont{D.}~\bibnamefont{Bonn}},
  \bibinfo{author}{\bibfnamefont{W.}~\bibnamefont{Hardy}},
  \bibnamefont{et~al.}, \bibinfo{journal}{Science}
  \textbf{\bibinfo{volume}{348}}, \bibinfo{pages}{317} (\bibinfo{year}{2015}).

\bibitem[{\citenamefont{Tranquada et~al.}(1994)\citenamefont{Tranquada,
  Buttrey, Sachan, and Lorenzo}}]{tranquada1994simultaneous}
\bibinfo{author}{\bibfnamefont{J.}~\bibnamefont{Tranquada}},
  \bibinfo{author}{\bibfnamefont{D.}~\bibnamefont{Buttrey}},
  \bibinfo{author}{\bibfnamefont{V.}~\bibnamefont{Sachan}}, \bibnamefont{and}
  \bibinfo{author}{\bibfnamefont{J.}~\bibnamefont{Lorenzo}},
  \bibinfo{journal}{Physical review letters} \textbf{\bibinfo{volume}{73}},
  \bibinfo{pages}{1003} (\bibinfo{year}{1994}).

\bibitem[{\citenamefont{Tranquada et~al.}(1995)\citenamefont{Tranquada,
  Sternlieb, Axe, Nakamura, and Uchida}}]{tranquada1995evidence}
\bibinfo{author}{\bibfnamefont{J.}~\bibnamefont{Tranquada}},
  \bibinfo{author}{\bibfnamefont{B.}~\bibnamefont{Sternlieb}},
  \bibinfo{author}{\bibfnamefont{J.}~\bibnamefont{Axe}},
  \bibinfo{author}{\bibfnamefont{Y.}~\bibnamefont{Nakamura}}, \bibnamefont{and}
  \bibinfo{author}{\bibfnamefont{S.}~\bibnamefont{Uchida}}
  (\bibinfo{year}{1995}).

\bibitem[{\citenamefont{Fujita et~al.}(2004)\citenamefont{Fujita, Goka, Yamada,
  Tranquada, and Regnault}}]{fujita2004stripe}
\bibinfo{author}{\bibfnamefont{M.}~\bibnamefont{Fujita}},
  \bibinfo{author}{\bibfnamefont{H.}~\bibnamefont{Goka}},
  \bibinfo{author}{\bibfnamefont{K.}~\bibnamefont{Yamada}},
  \bibinfo{author}{\bibfnamefont{J.}~\bibnamefont{Tranquada}},
  \bibnamefont{and} \bibinfo{author}{\bibfnamefont{L.}~\bibnamefont{Regnault}},
  \bibinfo{journal}{Physical Review B} \textbf{\bibinfo{volume}{70}},
  \bibinfo{pages}{104517} (\bibinfo{year}{2004}).

\bibitem[{\citenamefont{Howald et~al.}(2003)\citenamefont{Howald, Eisaki,
  Kaneko, Greven, and Kapitulnik}}]{howald2003periodic}
\bibinfo{author}{\bibfnamefont{C.}~\bibnamefont{Howald}},
  \bibinfo{author}{\bibfnamefont{H.}~\bibnamefont{Eisaki}},
  \bibinfo{author}{\bibfnamefont{N.}~\bibnamefont{Kaneko}},
  \bibinfo{author}{\bibfnamefont{M.}~\bibnamefont{Greven}}, \bibnamefont{and}
  \bibinfo{author}{\bibfnamefont{A.}~\bibnamefont{Kapitulnik}},
  \bibinfo{journal}{Physical Review B} \textbf{\bibinfo{volume}{67}},
  \bibinfo{pages}{014533} (\bibinfo{year}{2003}).

\bibitem[{\citenamefont{Hoffman et~al.}(2002)\citenamefont{Hoffman, Hudson,
  Lang, Madhavan, Eisaki, Uchida, and Davis}}]{hoffman2002four}
\bibinfo{author}{\bibfnamefont{J.}~\bibnamefont{Hoffman}},
  \bibinfo{author}{\bibfnamefont{E.}~\bibnamefont{Hudson}},
  \bibinfo{author}{\bibfnamefont{K.}~\bibnamefont{Lang}},
  \bibinfo{author}{\bibfnamefont{V.}~\bibnamefont{Madhavan}},
  \bibinfo{author}{\bibfnamefont{H.}~\bibnamefont{Eisaki}},
  \bibinfo{author}{\bibfnamefont{S.}~\bibnamefont{Uchida}}, \bibnamefont{and}
  \bibinfo{author}{\bibfnamefont{J.}~\bibnamefont{Davis}},
  \bibinfo{journal}{Science} \textbf{\bibinfo{volume}{295}},
  \bibinfo{pages}{466} (\bibinfo{year}{2002}).

\bibitem[{\citenamefont{Hanaguri et~al.}(2004)\citenamefont{Hanaguri, Lupien,
  Kohsaka, Lee, Azuma, Takano, Takagi, and Davis}}]{hanaguri2004checkerboard}
\bibinfo{author}{\bibfnamefont{T.}~\bibnamefont{Hanaguri}},
  \bibinfo{author}{\bibfnamefont{C.}~\bibnamefont{Lupien}},
  \bibinfo{author}{\bibfnamefont{Y.}~\bibnamefont{Kohsaka}},
  \bibinfo{author}{\bibfnamefont{D.-H.} \bibnamefont{Lee}},
  \bibinfo{author}{\bibfnamefont{M.}~\bibnamefont{Azuma}},
  \bibinfo{author}{\bibfnamefont{M.}~\bibnamefont{Takano}},
  \bibinfo{author}{\bibfnamefont{H.}~\bibnamefont{Takagi}}, \bibnamefont{and}
  \bibinfo{author}{\bibfnamefont{J.}~\bibnamefont{Davis}},
  \bibinfo{journal}{Nature} \textbf{\bibinfo{volume}{430}},
  \bibinfo{pages}{1001} (\bibinfo{year}{2004}).

\bibitem[{\citenamefont{Fink et~al.}(2009)\citenamefont{Fink, Schierle,
  Weschke, Geck, Hawthorn, Soltwisch, Wadati, Wu, D{\"u}rr, Wizent
  et~al.}}]{fink2009charge}
\bibinfo{author}{\bibfnamefont{J.}~\bibnamefont{Fink}},
  \bibinfo{author}{\bibfnamefont{E.}~\bibnamefont{Schierle}},
  \bibinfo{author}{\bibfnamefont{E.}~\bibnamefont{Weschke}},
  \bibinfo{author}{\bibfnamefont{J.}~\bibnamefont{Geck}},
  \bibinfo{author}{\bibfnamefont{D.}~\bibnamefont{Hawthorn}},
  \bibinfo{author}{\bibfnamefont{V.}~\bibnamefont{Soltwisch}},
  \bibinfo{author}{\bibfnamefont{H.}~\bibnamefont{Wadati}},
  \bibinfo{author}{\bibfnamefont{H.-H.} \bibnamefont{Wu}},
  \bibinfo{author}{\bibfnamefont{H.~A.} \bibnamefont{D{\"u}rr}},
  \bibinfo{author}{\bibfnamefont{N.}~\bibnamefont{Wizent}},
  \bibnamefont{et~al.}, \bibinfo{journal}{Physical Review B}
  \textbf{\bibinfo{volume}{79}}, \bibinfo{pages}{100502}
  (\bibinfo{year}{2009}).

\bibitem[{\citenamefont{Lalibert{\'e} et~al.}(2011)\citenamefont{Lalibert{\'e},
  Chang, Doiron-Leyraud, Hassinger, Daou, Rondeau, Ramshaw, Liang, Bonn, Hardy
  et~al.}}]{laliberte2011fermi}
\bibinfo{author}{\bibfnamefont{F.}~\bibnamefont{Lalibert{\'e}}},
  \bibinfo{author}{\bibfnamefont{J.}~\bibnamefont{Chang}},
  \bibinfo{author}{\bibfnamefont{N.}~\bibnamefont{Doiron-Leyraud}},
  \bibinfo{author}{\bibfnamefont{E.}~\bibnamefont{Hassinger}},
  \bibinfo{author}{\bibfnamefont{R.}~\bibnamefont{Daou}},
  \bibinfo{author}{\bibfnamefont{M.}~\bibnamefont{Rondeau}},
  \bibinfo{author}{\bibfnamefont{B.}~\bibnamefont{Ramshaw}},
  \bibinfo{author}{\bibfnamefont{R.}~\bibnamefont{Liang}},
  \bibinfo{author}{\bibfnamefont{D.}~\bibnamefont{Bonn}},
  \bibinfo{author}{\bibfnamefont{W.}~\bibnamefont{Hardy}},
  \bibnamefont{et~al.}, \bibinfo{journal}{Nature communications}
  \textbf{\bibinfo{volume}{2}}, \bibinfo{pages}{432} (\bibinfo{year}{2011}).

\bibitem[{\citenamefont{Chang et~al.}(2012)\citenamefont{Chang, Blackburn,
  Holmes, Christensen, Larsen, Mesot, Liang, Bonn, Hardy, Watenphul
  et~al.}}]{chang2012direct}
\bibinfo{author}{\bibfnamefont{J.}~\bibnamefont{Chang}},
  \bibinfo{author}{\bibfnamefont{E.}~\bibnamefont{Blackburn}},
  \bibinfo{author}{\bibfnamefont{A.}~\bibnamefont{Holmes}},
  \bibinfo{author}{\bibfnamefont{N.}~\bibnamefont{Christensen}},
  \bibinfo{author}{\bibfnamefont{J.}~\bibnamefont{Larsen}},
  \bibinfo{author}{\bibfnamefont{J.}~\bibnamefont{Mesot}},
  \bibinfo{author}{\bibfnamefont{R.}~\bibnamefont{Liang}},
  \bibinfo{author}{\bibfnamefont{D.}~\bibnamefont{Bonn}},
  \bibinfo{author}{\bibfnamefont{W.}~\bibnamefont{Hardy}},
  \bibinfo{author}{\bibfnamefont{A.}~\bibnamefont{Watenphul}},
  \bibnamefont{et~al.}, \bibinfo{journal}{Nature Physics}
  \textbf{\bibinfo{volume}{8}}, \bibinfo{pages}{871} (\bibinfo{year}{2012}).

\bibitem[{\citenamefont{Ghiringhelli et~al.}(2012)\citenamefont{Ghiringhelli,
  Le~Tacon, Minola, Blanco-Canosa, Mazzoli, Brookes, De~Luca, Frano, Hawthorn,
  He et~al.}}]{ghiringhelli2012long}
\bibinfo{author}{\bibfnamefont{G.}~\bibnamefont{Ghiringhelli}},
  \bibinfo{author}{\bibfnamefont{M.}~\bibnamefont{Le~Tacon}},
  \bibinfo{author}{\bibfnamefont{M.}~\bibnamefont{Minola}},
  \bibinfo{author}{\bibfnamefont{S.}~\bibnamefont{Blanco-Canosa}},
  \bibinfo{author}{\bibfnamefont{C.}~\bibnamefont{Mazzoli}},
  \bibinfo{author}{\bibfnamefont{N.}~\bibnamefont{Brookes}},
  \bibinfo{author}{\bibfnamefont{G.}~\bibnamefont{De~Luca}},
  \bibinfo{author}{\bibfnamefont{A.}~\bibnamefont{Frano}},
  \bibinfo{author}{\bibfnamefont{D.}~\bibnamefont{Hawthorn}},
  \bibinfo{author}{\bibfnamefont{F.}~\bibnamefont{He}}, \bibnamefont{et~al.},
  \bibinfo{journal}{Science} \textbf{\bibinfo{volume}{337}},
  \bibinfo{pages}{821} (\bibinfo{year}{2012}).

\bibitem[{\citenamefont{Achkar et~al.}(2012)\citenamefont{Achkar, Sutarto, Mao,
  He, Frano, Blanco-Canosa, Le~Tacon, Ghiringhelli, Braicovich, Minola
  et~al.}}]{RexsYBCO}
\bibinfo{author}{\bibfnamefont{A.~J.} \bibnamefont{Achkar}},
  \bibinfo{author}{\bibfnamefont{R.}~\bibnamefont{Sutarto}},
  \bibinfo{author}{\bibfnamefont{X.}~\bibnamefont{Mao}},
  \bibinfo{author}{\bibfnamefont{F.}~\bibnamefont{He}},
  \bibinfo{author}{\bibfnamefont{A.}~\bibnamefont{Frano}},
  \bibinfo{author}{\bibfnamefont{S.}~\bibnamefont{Blanco-Canosa}},
  \bibinfo{author}{\bibfnamefont{M.}~\bibnamefont{Le~Tacon}},
  \bibinfo{author}{\bibfnamefont{G.}~\bibnamefont{Ghiringhelli}},
  \bibinfo{author}{\bibfnamefont{L.}~\bibnamefont{Braicovich}},
  \bibinfo{author}{\bibfnamefont{M.}~\bibnamefont{Minola}},
  \bibnamefont{et~al.}, \bibinfo{journal}{Phys. Rev. Lett.}
  \textbf{\bibinfo{volume}{109}}, \bibinfo{pages}{167001}
  (\bibinfo{year}{2012}),
  \urlprefix\url{http://link.aps.org/doi/10.1103/PhysRevLett.109.167001}.

\bibitem[{\citenamefont{Doiron-Leyraud
  et~al.}(2013)\citenamefont{Doiron-Leyraud, Lepault, Cyr-Choiniere, Vignolle,
  Grissonnanche, Lalibert{\'e}, Chang, Bari{\v{s}}i{\'c}, Chan, Ji
  et~al.}}]{doiron2013hall}
\bibinfo{author}{\bibfnamefont{N.}~\bibnamefont{Doiron-Leyraud}},
  \bibinfo{author}{\bibfnamefont{S.}~\bibnamefont{Lepault}},
  \bibinfo{author}{\bibfnamefont{O.}~\bibnamefont{Cyr-Choiniere}},
  \bibinfo{author}{\bibfnamefont{B.}~\bibnamefont{Vignolle}},
  \bibinfo{author}{\bibfnamefont{G.}~\bibnamefont{Grissonnanche}},
  \bibinfo{author}{\bibfnamefont{F.}~\bibnamefont{Lalibert{\'e}}},
  \bibinfo{author}{\bibfnamefont{J.}~\bibnamefont{Chang}},
  \bibinfo{author}{\bibfnamefont{N.}~\bibnamefont{Bari{\v{s}}i{\'c}}},
  \bibinfo{author}{\bibfnamefont{M.}~\bibnamefont{Chan}},
  \bibinfo{author}{\bibfnamefont{L.}~\bibnamefont{Ji}}, \bibnamefont{et~al.},
  \bibinfo{journal}{Physical Review X} \textbf{\bibinfo{volume}{3}},
  \bibinfo{pages}{021019} (\bibinfo{year}{2013}).

\bibitem[{\citenamefont{LeBoeuf et~al.}(2013)\citenamefont{LeBoeuf, Kr{\"a}mer,
  Hardy, Liang, Bonn, and Proust}}]{leboeuf2013thermodynamic}
\bibinfo{author}{\bibfnamefont{D.}~\bibnamefont{LeBoeuf}},
  \bibinfo{author}{\bibfnamefont{S.}~\bibnamefont{Kr{\"a}mer}},
  \bibinfo{author}{\bibfnamefont{W.}~\bibnamefont{Hardy}},
  \bibinfo{author}{\bibfnamefont{R.}~\bibnamefont{Liang}},
  \bibinfo{author}{\bibfnamefont{D.}~\bibnamefont{Bonn}}, \bibnamefont{and}
  \bibinfo{author}{\bibfnamefont{C.}~\bibnamefont{Proust}},
  \bibinfo{journal}{Nature Physics} \textbf{\bibinfo{volume}{9}},
  \bibinfo{pages}{79} (\bibinfo{year}{2013}).

\bibitem[{\citenamefont{He et~al.}(2014)\citenamefont{He, Yin, Zech,
  Soumyanarayanan, Yee, Williams, Boyer, Chatterjee, Wise, Zeljkovic
  et~al.}}]{he2014fermi}
\bibinfo{author}{\bibfnamefont{Y.}~\bibnamefont{He}},
  \bibinfo{author}{\bibfnamefont{Y.}~\bibnamefont{Yin}},
  \bibinfo{author}{\bibfnamefont{M.}~\bibnamefont{Zech}},
  \bibinfo{author}{\bibfnamefont{A.}~\bibnamefont{Soumyanarayanan}},
  \bibinfo{author}{\bibfnamefont{M.~M.} \bibnamefont{Yee}},
  \bibinfo{author}{\bibfnamefont{T.}~\bibnamefont{Williams}},
  \bibinfo{author}{\bibfnamefont{M.}~\bibnamefont{Boyer}},
  \bibinfo{author}{\bibfnamefont{K.}~\bibnamefont{Chatterjee}},
  \bibinfo{author}{\bibfnamefont{W.}~\bibnamefont{Wise}},
  \bibinfo{author}{\bibfnamefont{I.}~\bibnamefont{Zeljkovic}},
  \bibnamefont{et~al.}, \bibinfo{journal}{Science}
  \textbf{\bibinfo{volume}{344}}, \bibinfo{pages}{608} (\bibinfo{year}{2014}).

\bibitem[{\citenamefont{Tabis et~al.}(2014)\citenamefont{Tabis, Li, Le~Tacon,
  Braicovich, Kreyssig, Minola, Dellea, Weschke, Veit, Ramazanoglu
  et~al.}}]{tabis2014charge}
\bibinfo{author}{\bibfnamefont{W.}~\bibnamefont{Tabis}},
  \bibinfo{author}{\bibfnamefont{Y.}~\bibnamefont{Li}},
  \bibinfo{author}{\bibfnamefont{M.}~\bibnamefont{Le~Tacon}},
  \bibinfo{author}{\bibfnamefont{L.}~\bibnamefont{Braicovich}},
  \bibinfo{author}{\bibfnamefont{A.}~\bibnamefont{Kreyssig}},
  \bibinfo{author}{\bibfnamefont{M.}~\bibnamefont{Minola}},
  \bibinfo{author}{\bibfnamefont{G.}~\bibnamefont{Dellea}},
  \bibinfo{author}{\bibfnamefont{E.}~\bibnamefont{Weschke}},
  \bibinfo{author}{\bibfnamefont{M.}~\bibnamefont{Veit}},
  \bibinfo{author}{\bibfnamefont{M.}~\bibnamefont{Ramazanoglu}},
  \bibnamefont{et~al.}, \bibinfo{journal}{Nature communications}
  \textbf{\bibinfo{volume}{5}} (\bibinfo{year}{2014}).

\bibitem[{\citenamefont{Forgan et~al.}(2015)\citenamefont{Forgan, Blackburn,
  Holmes, Briffa, Chang, Bouchenoire, Brown, Liang, Bonn, Hardy
  et~al.}}]{forgan2015nature}
\bibinfo{author}{\bibfnamefont{E.}~\bibnamefont{Forgan}},
  \bibinfo{author}{\bibfnamefont{E.}~\bibnamefont{Blackburn}},
  \bibinfo{author}{\bibfnamefont{A.}~\bibnamefont{Holmes}},
  \bibinfo{author}{\bibfnamefont{A.}~\bibnamefont{Briffa}},
  \bibinfo{author}{\bibfnamefont{J.}~\bibnamefont{Chang}},
  \bibinfo{author}{\bibfnamefont{L.}~\bibnamefont{Bouchenoire}},
  \bibinfo{author}{\bibfnamefont{S.}~\bibnamefont{Brown}},
  \bibinfo{author}{\bibfnamefont{R.}~\bibnamefont{Liang}},
  \bibinfo{author}{\bibfnamefont{D.}~\bibnamefont{Bonn}},
  \bibinfo{author}{\bibfnamefont{W.}~\bibnamefont{Hardy}},
  \bibnamefont{et~al.}, \bibinfo{journal}{arXiv preprint arXiv:1504.01585}
  (\bibinfo{year}{2015}).

\bibitem[{\citenamefont{Chakravarty et~al.}(2001)\citenamefont{Chakravarty,
  Laughlin, Morr, and Nayak}}]{Chakravarty2001}
\bibinfo{author}{\bibfnamefont{S.}~\bibnamefont{Chakravarty}},
  \bibinfo{author}{\bibfnamefont{R.~B.} \bibnamefont{Laughlin}},
  \bibinfo{author}{\bibfnamefont{D.~K.} \bibnamefont{Morr}}, \bibnamefont{and}
  \bibinfo{author}{\bibfnamefont{C.}~\bibnamefont{Nayak}},
  \bibinfo{journal}{Phys. Rev. B} \textbf{\bibinfo{volume}{63}},
  \bibinfo{pages}{094503} (\bibinfo{year}{2001}),
  \urlprefix\url{http://link.aps.org/doi/10.1103/PhysRevB.63.094503}.

\bibitem[{\citenamefont{Millis and Norman}(2007)}]{millis2007antiphase}
\bibinfo{author}{\bibfnamefont{A.~J.} \bibnamefont{Millis}} \bibnamefont{and}
  \bibinfo{author}{\bibfnamefont{M.}~\bibnamefont{Norman}},
  \bibinfo{journal}{Physical Review B} \textbf{\bibinfo{volume}{76}},
  \bibinfo{pages}{220503} (\bibinfo{year}{2007}).

\bibitem[{\citenamefont{Harrison and Sebastian}(2011)}]{harrison2011protected}
\bibinfo{author}{\bibfnamefont{N.}~\bibnamefont{Harrison}} \bibnamefont{and}
  \bibinfo{author}{\bibfnamefont{S.}~\bibnamefont{Sebastian}},
  \bibinfo{journal}{Physical Review Letters} \textbf{\bibinfo{volume}{106}},
  \bibinfo{pages}{226402} (\bibinfo{year}{2011}).

\bibitem[{\citenamefont{Eun et~al.}(2012)\citenamefont{Eun, Wang, and
  Chakravarty}}]{Eun2012}
\bibinfo{author}{\bibfnamefont{J.}~\bibnamefont{Eun}},
  \bibinfo{author}{\bibfnamefont{Z.}~\bibnamefont{Wang}}, \bibnamefont{and}
  \bibinfo{author}{\bibfnamefont{S.}~\bibnamefont{Chakravarty}},
  \bibinfo{journal}{Proceedings of the National Academy of Sciences}
  \textbf{\bibinfo{volume}{109}}, \bibinfo{pages}{13198}
  (\bibinfo{year}{2012}),
  \eprint{http://www.pnas.org/content/109/33/13198.full.pdf+html},
  \urlprefix\url{http://www.pnas.org/content/109/33/13198.abstract}.

\bibitem[{\citenamefont{Harrison and Sebastian}(2012)}]{harrison2012fermi}
\bibinfo{author}{\bibfnamefont{N.}~\bibnamefont{Harrison}} \bibnamefont{and}
  \bibinfo{author}{\bibfnamefont{S.}~\bibnamefont{Sebastian}},
  \bibinfo{journal}{New Journal of Physics} \textbf{\bibinfo{volume}{14}},
  \bibinfo{pages}{095023} (\bibinfo{year}{2012}).

\bibitem[{\citenamefont{Lee}(2014)}]{lee2014amperean}
\bibinfo{author}{\bibfnamefont{P.~A.} \bibnamefont{Lee}},
  \bibinfo{journal}{Physical Review X} \textbf{\bibinfo{volume}{4}},
  \bibinfo{pages}{031017} (\bibinfo{year}{2014}).

\bibitem[{\citenamefont{Allais et~al.}(2014)\citenamefont{Allais, Chowdhury,
  and Sachdev}}]{allais2014connecting}
\bibinfo{author}{\bibfnamefont{A.}~\bibnamefont{Allais}},
  \bibinfo{author}{\bibfnamefont{D.}~\bibnamefont{Chowdhury}},
  \bibnamefont{and} \bibinfo{author}{\bibfnamefont{S.}~\bibnamefont{Sachdev}},
  \bibinfo{journal}{Nature communications} \textbf{\bibinfo{volume}{5}}
  (\bibinfo{year}{2014}).

\bibitem[{\citenamefont{Wang and Chubukov}(2014)}]{wang2014charge}
\bibinfo{author}{\bibfnamefont{Y.}~\bibnamefont{Wang}} \bibnamefont{and}
  \bibinfo{author}{\bibfnamefont{A.}~\bibnamefont{Chubukov}},
  \bibinfo{journal}{Physical Review B} \textbf{\bibinfo{volume}{90}},
  \bibinfo{pages}{035149} (\bibinfo{year}{2014}).

\bibitem[{\citenamefont{Maharaj et~al.}(2014)\citenamefont{Maharaj, Hosur, and
  Raghu}}]{Maharaj2014}
\bibinfo{author}{\bibfnamefont{A.~V.} \bibnamefont{Maharaj}},
  \bibinfo{author}{\bibfnamefont{P.}~\bibnamefont{Hosur}}, \bibnamefont{and}
  \bibinfo{author}{\bibfnamefont{S.}~\bibnamefont{Raghu}},
  \bibinfo{journal}{Phys. Rev. B} \textbf{\bibinfo{volume}{90}},
  \bibinfo{pages}{125108} (\bibinfo{year}{2014}),
  \urlprefix\url{http://link.aps.org/doi/10.1103/PhysRevB.90.125108}.

\bibitem[{\citenamefont{Russo and Chakravarty}(2015)}]{russo2015random}
\bibinfo{author}{\bibfnamefont{A.}~\bibnamefont{Russo}} \bibnamefont{and}
  \bibinfo{author}{\bibfnamefont{S.}~\bibnamefont{Chakravarty}},
  \bibinfo{journal}{arXiv preprint arXiv:1504.03378}  (\bibinfo{year}{2015}).

\bibitem[{\citenamefont{Briffa et~al.}(2015)\citenamefont{Briffa, Blackburn,
  Hayden, Yelland, M.W., and Forgan}}]{briffa2015fermi}
\bibinfo{author}{\bibfnamefont{A.}~\bibnamefont{Briffa}},
  \bibinfo{author}{\bibfnamefont{E.}~\bibnamefont{Blackburn}},
  \bibinfo{author}{\bibfnamefont{S.}~\bibnamefont{Hayden}},
  \bibinfo{author}{\bibfnamefont{E.}~\bibnamefont{Yelland}},
  \bibinfo{author}{\bibfnamefont{L.}~\bibnamefont{M.W.}}, \bibnamefont{and}
  \bibinfo{author}{\bibfnamefont{E.}~\bibnamefont{Forgan}},
  \bibinfo{journal}{arXiv preprint arXiv:1510.02603}  (\bibinfo{year}{2015}).

\bibitem[{\citenamefont{Harrison et~al.}(2015)\citenamefont{Harrison, Ramshaw,
  and Shekhter}}]{harrison2015nodal}
\bibinfo{author}{\bibfnamefont{N.}~\bibnamefont{Harrison}},
  \bibinfo{author}{\bibfnamefont{B.}~\bibnamefont{Ramshaw}}, \bibnamefont{and}
  \bibinfo{author}{\bibfnamefont{A.}~\bibnamefont{Shekhter}},
  \bibinfo{journal}{Scientific reports} \textbf{\bibinfo{volume}{5}}
  (\bibinfo{year}{2015}).

\bibitem[{\citenamefont{Yao et~al.}(2011)\citenamefont{Yao, Lee, and
  Kivelson}}]{yao2011fermi}
\bibinfo{author}{\bibfnamefont{H.}~\bibnamefont{Yao}},
  \bibinfo{author}{\bibfnamefont{D.-H.} \bibnamefont{Lee}}, \bibnamefont{and}
  \bibinfo{author}{\bibfnamefont{S.}~\bibnamefont{Kivelson}},
  \bibinfo{journal}{Physical Review B} \textbf{\bibinfo{volume}{84}},
  \bibinfo{pages}{012507} (\bibinfo{year}{2011}).

\bibitem[{\citenamefont{Riggs et~al.}(2011)\citenamefont{Riggs, Vafek, Kemper,
  Betts, Migliori, Balakirev, Hardy, Liang, Bonn, and
  Boebinger}}]{riggs2011heat}
\bibinfo{author}{\bibfnamefont{S.~C.} \bibnamefont{Riggs}},
  \bibinfo{author}{\bibfnamefont{O.}~\bibnamefont{Vafek}},
  \bibinfo{author}{\bibfnamefont{J.}~\bibnamefont{Kemper}},
  \bibinfo{author}{\bibfnamefont{J.}~\bibnamefont{Betts}},
  \bibinfo{author}{\bibfnamefont{A.}~\bibnamefont{Migliori}},
  \bibinfo{author}{\bibfnamefont{F.}~\bibnamefont{Balakirev}},
  \bibinfo{author}{\bibfnamefont{W.}~\bibnamefont{Hardy}},
  \bibinfo{author}{\bibfnamefont{R.}~\bibnamefont{Liang}},
  \bibinfo{author}{\bibfnamefont{D.}~\bibnamefont{Bonn}}, \bibnamefont{and}
  \bibinfo{author}{\bibfnamefont{G.}~\bibnamefont{Boebinger}},
  \bibinfo{journal}{Nature Physics} \textbf{\bibinfo{volume}{7}},
  \bibinfo{pages}{332} (\bibinfo{year}{2011}).

\bibitem[{\citenamefont{Doiron-Leyraud
  et~al.}(2015)\citenamefont{Doiron-Leyraud, Badoux, de~Cotret, Lepault,
  LeBoeuf, Lalibert{\'e}, Hassinger, Ramshaw, Bonn, Hardy
  et~al.}}]{doiron2015evidence}
\bibinfo{author}{\bibfnamefont{N.}~\bibnamefont{Doiron-Leyraud}},
  \bibinfo{author}{\bibfnamefont{S.}~\bibnamefont{Badoux}},
  \bibinfo{author}{\bibfnamefont{S.~R.} \bibnamefont{de~Cotret}},
  \bibinfo{author}{\bibfnamefont{S.}~\bibnamefont{Lepault}},
  \bibinfo{author}{\bibfnamefont{D.}~\bibnamefont{LeBoeuf}},
  \bibinfo{author}{\bibfnamefont{F.}~\bibnamefont{Lalibert{\'e}}},
  \bibinfo{author}{\bibfnamefont{E.}~\bibnamefont{Hassinger}},
  \bibinfo{author}{\bibfnamefont{B.}~\bibnamefont{Ramshaw}},
  \bibinfo{author}{\bibfnamefont{D.}~\bibnamefont{Bonn}},
  \bibinfo{author}{\bibfnamefont{W.}~\bibnamefont{Hardy}},
  \bibnamefont{et~al.}, \bibinfo{journal}{Nature communications}
  \textbf{\bibinfo{volume}{6}} (\bibinfo{year}{2015}).

\bibitem[{\citenamefont{Wang and Chakravarty}(2015)}]{wang2015onsager}
\bibinfo{author}{\bibfnamefont{Z.}~\bibnamefont{Wang}} \bibnamefont{and}
  \bibinfo{author}{\bibfnamefont{S.}~\bibnamefont{Chakravarty}},
  \bibinfo{journal}{arXiv preprint arXiv:1509.00494}  (\bibinfo{year}{2015}).

\bibitem[{\citenamefont{Andersen et~al.}(1995)\citenamefont{Andersen,
  Liechtenstein, Jepsen, and Paulsen}}]{andersen1995lda}
\bibinfo{author}{\bibfnamefont{O.}~\bibnamefont{Andersen}},
  \bibinfo{author}{\bibfnamefont{A.}~\bibnamefont{Liechtenstein}},
  \bibinfo{author}{\bibfnamefont{O.}~\bibnamefont{Jepsen}}, \bibnamefont{and}
  \bibinfo{author}{\bibfnamefont{F.}~\bibnamefont{Paulsen}},
  \bibinfo{journal}{Journal of Physics and Chemistry of Solids}
  \textbf{\bibinfo{volume}{56}}, \bibinfo{pages}{1573} (\bibinfo{year}{1995}).

\bibitem[{\citenamefont{Elfimov et~al.}(2008)\citenamefont{Elfimov, Sawatzky,
  and Damascelli}}]{elfimov2008theory}
\bibinfo{author}{\bibfnamefont{I.}~\bibnamefont{Elfimov}},
  \bibinfo{author}{\bibfnamefont{G.}~\bibnamefont{Sawatzky}}, \bibnamefont{and}
  \bibinfo{author}{\bibfnamefont{A.}~\bibnamefont{Damascelli}},
  \bibinfo{journal}{Physical Review B} \textbf{\bibinfo{volume}{77}},
  \bibinfo{pages}{060504} (\bibinfo{year}{2008}).

\bibitem[{\citenamefont{Fournier et~al.}(2010)\citenamefont{Fournier, Levy,
  Pennec, McChesney, Bostwick, Rotenberg, Liang, Hardy, Bonn, Elfimov
  et~al.}}]{fournier2010loss}
\bibinfo{author}{\bibfnamefont{D.}~\bibnamefont{Fournier}},
  \bibinfo{author}{\bibfnamefont{G.}~\bibnamefont{Levy}},
  \bibinfo{author}{\bibfnamefont{Y.}~\bibnamefont{Pennec}},
  \bibinfo{author}{\bibfnamefont{J.}~\bibnamefont{McChesney}},
  \bibinfo{author}{\bibfnamefont{A.}~\bibnamefont{Bostwick}},
  \bibinfo{author}{\bibfnamefont{E.}~\bibnamefont{Rotenberg}},
  \bibinfo{author}{\bibfnamefont{R.}~\bibnamefont{Liang}},
  \bibinfo{author}{\bibfnamefont{W.}~\bibnamefont{Hardy}},
  \bibinfo{author}{\bibfnamefont{D.}~\bibnamefont{Bonn}},
  \bibinfo{author}{\bibfnamefont{I.}~\bibnamefont{Elfimov}},
  \bibnamefont{et~al.}, \bibinfo{journal}{Nature Physics}
  \textbf{\bibinfo{volume}{6}}, \bibinfo{pages}{905} (\bibinfo{year}{2010}).

\bibitem[{\citenamefont{Chakravarty et~al.}(1999)\citenamefont{Chakravarty,
  Kee, and Abrahams}}]{chakravarty1999frustrated}
\bibinfo{author}{\bibfnamefont{S.}~\bibnamefont{Chakravarty}},
  \bibinfo{author}{\bibfnamefont{H.-Y.} \bibnamefont{Kee}}, \bibnamefont{and}
  \bibinfo{author}{\bibfnamefont{E.}~\bibnamefont{Abrahams}},
  \bibinfo{journal}{Physical review letters} \textbf{\bibinfo{volume}{82}},
  \bibinfo{pages}{2366} (\bibinfo{year}{1999}).

\bibitem[{\citenamefont{Ioffe and Millis}(1999)}]{ioffe1999superconductivity}
\bibinfo{author}{\bibfnamefont{L.}~\bibnamefont{Ioffe}} \bibnamefont{and}
  \bibinfo{author}{\bibfnamefont{A.}~\bibnamefont{Millis}},
  \bibinfo{journal}{Science} \textbf{\bibinfo{volume}{285}},
  \bibinfo{pages}{1241} (\bibinfo{year}{1999}).

\bibitem[{\citenamefont{Carlson et~al.}(2000)\citenamefont{Carlson, Orgad,
  Kivelson, and Emery}}]{carlson2000dimensional}
\bibinfo{author}{\bibfnamefont{E.}~\bibnamefont{Carlson}},
  \bibinfo{author}{\bibfnamefont{D.}~\bibnamefont{Orgad}},
  \bibinfo{author}{\bibfnamefont{S.}~\bibnamefont{Kivelson}}, \bibnamefont{and}
  \bibinfo{author}{\bibfnamefont{V.}~\bibnamefont{Emery}},
  \bibinfo{journal}{Physical Review B} \textbf{\bibinfo{volume}{62}},
  \bibinfo{pages}{3422} (\bibinfo{year}{2000}).

\bibitem[{\citenamefont{Zhang et~al.}(2015)\citenamefont{Zhang, Maharaj, and
  Kivelson}}]{zhang2015disruption}
\bibinfo{author}{\bibfnamefont{Y.}~\bibnamefont{Zhang}},
  \bibinfo{author}{\bibfnamefont{A.~V.} \bibnamefont{Maharaj}},
  \bibnamefont{and} \bibinfo{author}{\bibfnamefont{S.}~\bibnamefont{Kivelson}},
  \bibinfo{journal}{Physical Review B} \textbf{\bibinfo{volume}{91}},
  \bibinfo{pages}{085105} (\bibinfo{year}{2015}).

\bibitem[{\citenamefont{Bergemann et~al.}(2003)\citenamefont{Bergemann,
  Mackenzie, Julian, Forsythe, and Ohmichi}}]{Bergemann:2003}
\bibinfo{author}{\bibfnamefont{C.}~\bibnamefont{Bergemann}},
  \bibinfo{author}{\bibfnamefont{A.}~\bibnamefont{Mackenzie}},
  \bibinfo{author}{\bibfnamefont{S.}~\bibnamefont{Julian}},
  \bibinfo{author}{\bibfnamefont{D.}~\bibnamefont{Forsythe}}, \bibnamefont{and}
  \bibinfo{author}{\bibfnamefont{E.}~\bibnamefont{Ohmichi}},
  \bibinfo{journal}{Advances in Physics} \textbf{\bibinfo{volume}{52}},
  \bibinfo{pages}{639} (\bibinfo{year}{2003}), ISSN \bibinfo{issn}{0001-8732}.

\bibitem[{\citenamefont{Chakravarty et~al.}(1993)\citenamefont{Chakravarty,
  Sudb{\o}, Anderson, and Strong}}]{chakravarty1993interlayer}
\bibinfo{author}{\bibfnamefont{S.}~\bibnamefont{Chakravarty}},
  \bibinfo{author}{\bibfnamefont{A.}~\bibnamefont{Sudb{\o}}},
  \bibinfo{author}{\bibfnamefont{P.~W.} \bibnamefont{Anderson}},
  \bibnamefont{and} \bibinfo{author}{\bibfnamefont{S.}~\bibnamefont{Strong}},
  \bibinfo{journal}{Science} \textbf{\bibinfo{volume}{261}},
  \bibinfo{pages}{337} (\bibinfo{year}{1993}).

\bibitem[{\citenamefont{Gerber et~al.}(2015)\citenamefont{Gerber, Jang, Nojiri,
  Matsuzawa, Yasumura, Bonn, Liang, Hardy, Islam, Mehta
  et~al.}}]{gerber2015three}
\bibinfo{author}{\bibfnamefont{S.}~\bibnamefont{Gerber}},
  \bibinfo{author}{\bibfnamefont{H.}~\bibnamefont{Jang}},
  \bibinfo{author}{\bibfnamefont{H.}~\bibnamefont{Nojiri}},
  \bibinfo{author}{\bibfnamefont{S.}~\bibnamefont{Matsuzawa}},
  \bibinfo{author}{\bibfnamefont{H.}~\bibnamefont{Yasumura}},
  \bibinfo{author}{\bibfnamefont{D.}~\bibnamefont{Bonn}},
  \bibinfo{author}{\bibfnamefont{R.}~\bibnamefont{Liang}},
  \bibinfo{author}{\bibfnamefont{W.}~\bibnamefont{Hardy}},
  \bibinfo{author}{\bibfnamefont{Z.}~\bibnamefont{Islam}},
  \bibinfo{author}{\bibfnamefont{A.}~\bibnamefont{Mehta}},
  \bibnamefont{et~al.}, \bibinfo{journal}{arXiv preprint arXiv:1506.07910}
  (\bibinfo{year}{2015}).

\bibitem[{\citenamefont{Shoenberg}(1984)}]{shoenberg:1984}
\bibinfo{author}{\bibfnamefont{D.}~\bibnamefont{Shoenberg}},
  \emph{\bibinfo{title}{Magnetic oscillations in metals}}, Cambridge monographs
  on physics (\bibinfo{publisher}{Cambridge University Press},
  \bibinfo{year}{1984}), ISBN \bibinfo{isbn}{9780521224802}.

\bibitem[{\citenamefont{Comin et~al.}(2015)\citenamefont{Comin, Sutarto, He,
  da~Silva~Neto, Chauviere, Frano, Liang, Hardy, Bonn, Yoshida
  et~al.}}]{comin2015symmetry}
\bibinfo{author}{\bibfnamefont{R.}~\bibnamefont{Comin}},
  \bibinfo{author}{\bibfnamefont{R.}~\bibnamefont{Sutarto}},
  \bibinfo{author}{\bibfnamefont{F.}~\bibnamefont{He}},
  \bibinfo{author}{\bibfnamefont{E.}~\bibnamefont{da~Silva~Neto}},
  \bibinfo{author}{\bibfnamefont{L.}~\bibnamefont{Chauviere}},
  \bibinfo{author}{\bibfnamefont{A.}~\bibnamefont{Frano}},
  \bibinfo{author}{\bibfnamefont{R.}~\bibnamefont{Liang}},
  \bibinfo{author}{\bibfnamefont{W.}~\bibnamefont{Hardy}},
  \bibinfo{author}{\bibfnamefont{D.}~\bibnamefont{Bonn}},
  \bibinfo{author}{\bibfnamefont{Y.}~\bibnamefont{Yoshida}},
  \bibnamefont{et~al.}, \bibinfo{journal}{Nature materials}
  (\bibinfo{year}{2015}).

\bibitem[{\citenamefont{Chowdhury and Sachdev}(2015)}]{chowdhury2015higgs}
\bibinfo{author}{\bibfnamefont{D.}~\bibnamefont{Chowdhury}} \bibnamefont{and}
  \bibinfo{author}{\bibfnamefont{S.}~\bibnamefont{Sachdev}},
  \bibinfo{journal}{Physical Review B} \textbf{\bibinfo{volume}{91}},
  \bibinfo{pages}{115123} (\bibinfo{year}{2015}).

\bibitem[{\citenamefont{Chowdhury and Sachdev}(2014)}]{chowdhury2014Density}
\bibinfo{author}{\bibfnamefont{D.}~\bibnamefont{Chowdhury}} \bibnamefont{and}
  \bibinfo{author}{\bibfnamefont{S.}~\bibnamefont{Sachdev}},
  \bibinfo{journal}{Phys. Rev. B} \textbf{\bibinfo{volume}{90}},
  \bibinfo{pages}{245136} (\bibinfo{year}{2014}),
  \urlprefix\url{http://link.aps.org/doi/10.1103/PhysRevB.90.245136}.

\bibitem[{\citenamefont{Hlobil et~al.}(2015)\citenamefont{Hlobil, Maharaj,
  Hosur, Shapiro, Fisher, and Raghu}}]{hlobil2015elastoconductivity}
\bibinfo{author}{\bibfnamefont{P.}~\bibnamefont{Hlobil}},
  \bibinfo{author}{\bibfnamefont{A.~V.} \bibnamefont{Maharaj}},
  \bibinfo{author}{\bibfnamefont{P.}~\bibnamefont{Hosur}},
  \bibinfo{author}{\bibfnamefont{M.}~\bibnamefont{Shapiro}},
  \bibinfo{author}{\bibfnamefont{I.}~\bibnamefont{Fisher}}, \bibnamefont{and}
  \bibinfo{author}{\bibfnamefont{S.}~\bibnamefont{Raghu}},
  \bibinfo{journal}{Physical Review B} \textbf{\bibinfo{volume}{92}},
  \bibinfo{pages}{035148} (\bibinfo{year}{2015}).

\end{thebibliography}
\begin{widetext}
\appendix
\section{Form of the Hamiltonian for general angles $\phi$}
\label{app:technical}
For additional azimuthal angles as $\tan\phi=1/M$ where
$M\in\mathbb{Z}$ (or equivalently $\tan\phi=M$ by symmetry):
\begin{equation}
\vec{B}=B\left(\hat{z}\cos\theta+\hat{x}\sin\theta
M/\sqrt{M^{2}+1}+\hat{y}\sin\theta/\sqrt{M^{2}+1}\right)
\end{equation}
we can no longer keep the translation symmetry along the $\hat{y}$
direction for arbitrary $B$ with the chosen Landau gauge, however,
we can define the new magnetic unit cell with the new lattice
vectors $\hat{x}'=\hat{x},$ and $\hat{y}'=M\hat{x}+\hat{y}$ along
$\vec{B}$ in plane or equivalently $x'=x-My$, $y'=y$. Once again we
can choose a proper gauge so that the translation symmetry along the
$\hat{y}'$ direction is preserved:
\begin{align}
\bm{A} &= \left(0, 2\pi \Phi (x-My) , -2\pi \Phi  a_c \tan\theta
(x-My)/\sqrt{M^{2}+1}
\tan{\theta}\right) \nonumber\\
&= \left(0, 2\pi \Phi x' , -2\pi \Phi x'  a_c
\tan\theta/\sqrt{M^{2}+1}\right)
\end{align}
where $\Phi = B\cos\theta$ is the magnetic flux through the
plaquette in the $x-y$ plaquette, and $\Phi  a_c
\tan\theta/\sqrt{M^{2}+1} $ is the flux through the  $x-z$
plaquette. The hopping matrix elements no longer depend on $y'$,
therefore we can Fourier transform into the corresponding $k_{y}'$
momentum basis. The resulting Hamiltonian (for each $k_{y}'$ and
spin $\sigma$) becomes:
\begin{eqnarray}
\hat {H}_{k_{y}',\sigma}&=&
\underset{x,\lambda}{\sum}t_{x,\lambda}\left[c_{x+1,\lambda}^{\dagger}c_{x,\lambda}+\mbox{H.c.}\right]+\frac{4\pi
\tilde{g} \Phi \sigma}{\cos \theta}
c_{x,\lambda}^{\dagger}c_{x,\lambda}
\\\nonumber&+&\underset{x,\lambda}{\sum}t_{y,\lambda}\left[c_{x-M,\lambda}^{\dagger}c_{x,\lambda}\exp\left(i2\pi \Phi x-k_{y}'\right)+\mbox{H.c.}\right]
\\\nonumber&+&\underset{x}{\sum}t_{\perp}\left[c_{x,2}^{\dagger}c_{x,1}\exp\left(i 2\pi \Phi x a_c
\tan\theta/\sqrt{M^{2}+1}\right)+\mbox{H.c.}\right]
\end{eqnarray}
where we have suppressed the $k_{y}'$ and $\sigma$ labels in the
fermion operators. The Hamiltonian is still block tri-diagonal and
its physical properties including DOS can be efficiently calculated
using recursive Green's function method.

\section{Mirror symmetry and the absence of breakdown frequencies}
\label{sec:mirrorsymmetry}
\begin{figure}[t]
\begin{center}
\subfigure[ {Mirror symmetric}]{
\includegraphics[width=0.45\textwidth]{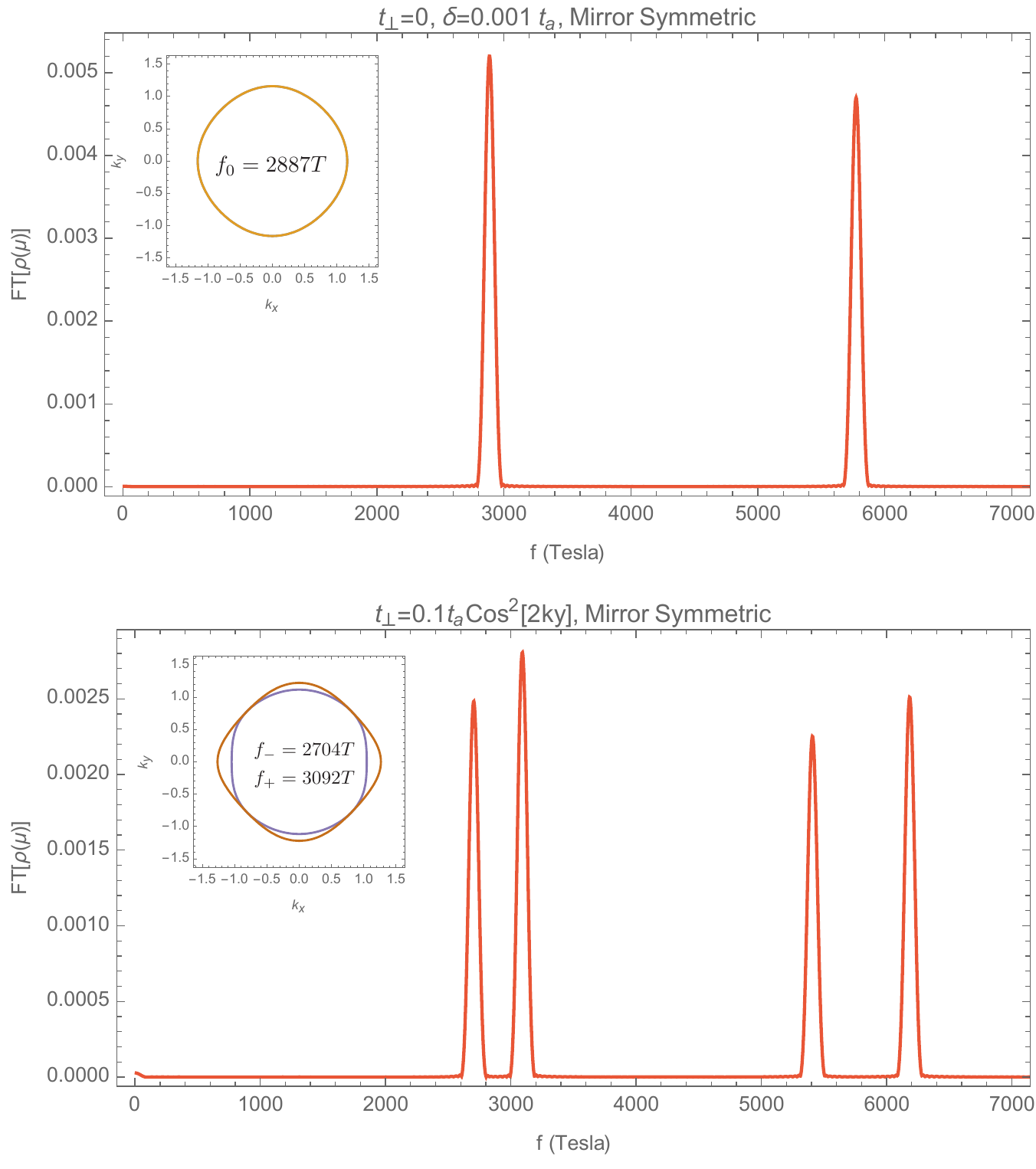}\label{fig:symmetricfs}}
\subfigure[ {Broken mirror symmetry}]{
\includegraphics[width=0.465\textwidth]{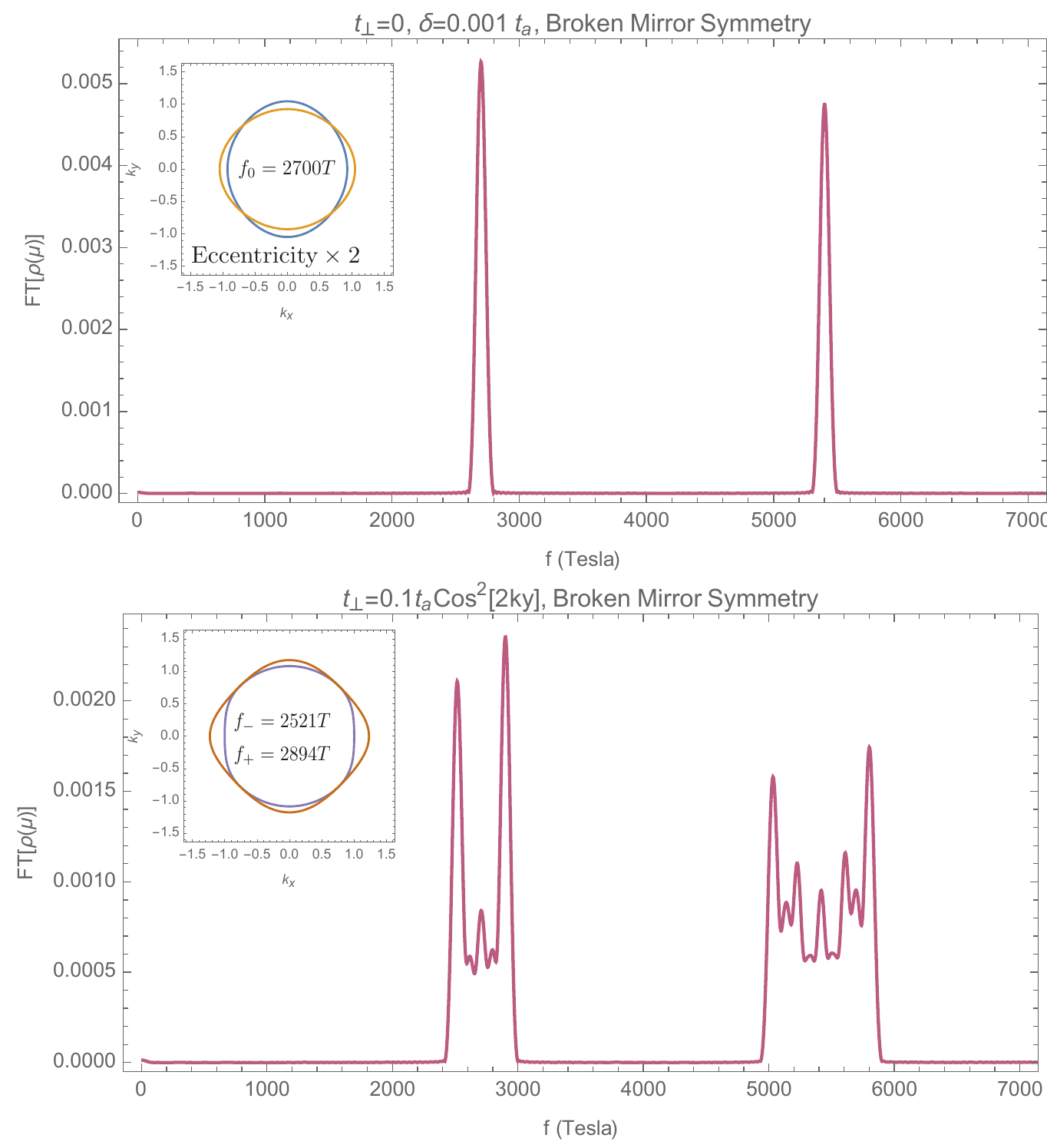}\label{fig:asymmetricfs}}
\caption{Fourier transforms of QOs in the density of states for two
models (a) with mirror symmetry and (b) without mirror symmetry. In
(a) we consider identical Fermi surfaces with an interlayer
tunneling of the form $t_{\perp}(\k) = t_{\perp}\cos^{2}(2k_y)$,
while in (b) the mirror symmetry is weakly broken by considering
orthogonal Fermi surfaces with a weak mass anisotropy ($t_b =
0.95t_a$) and the interlayer tunneling is once more $t_{\perp}(\k) =
t_{\perp}\cos^{2}(2k_y)$. The bilayer bonding and anti bonding Fermi
surfaces are almost identical in both cases, yet the QO frequencies
are dramatically different: mirror symmetry forbids breakdown orbits
in (a). }
\end{center}
\end{figure}
Here we discuss in further detail the absence of magnetic breakdown
when a mirror symmetry relating the two planes of the bilayer is
present. The essence of this symmetry argument is the following:  in
the presence of a magnetic field semiclassical dynamics correctly
captures the motion of electrons, while magnetic breakdown is
allowed as long as there exist matrix elements that take electrons
from one orbit to the next. However, if there is a mirror plane
perpendicular to the magnetic field, the mirror parity of the states
remains a good quantum number even in the presence of a magnetic
field.  There are necessarily no matrix elements between states with
different quantum numbers, and so breakdown processes are forbidden
by this symmetry. We emphasize that this argument is also applicable
in the limit of a single bilayer, i.e. when $k_z$ is not a good
quantum number.

This symmetry may be viewed at a more operational level by
considering the Hamiltonian of a bilayer with identical dispersions
$\varepsilon(\k)$ in each layer. In the absence of a field, this
takes the form
\begin{align}
H &= \sum_{\bm{\k}} \Psi_{\k} \hat{H}_{\bm{\k}}\Psi_{\k} = \sum_{\bm{k} = k_x,k_y}\begin{matrix}
 (c^{\dag}_{\bm{k},1}& c^{\dag}_{\bm{k},2}) \\
 &\\
 \end{matrix}
\begin{pmatrix}
\varepsilon(\bm{k})& t_{\perp}(\k)\\
t_{\perp}(\k) & \varepsilon(\bm{k})
\end{pmatrix}
\begin{pmatrix}
c_{\bm{k},1}\\
c_{\bm{k},2}
\end{pmatrix}
\end{align}
where $t_{\perp}(\k)$ is the (in general) momentum dependent tunneling between layers.

Mirror symmetry relating the two layers of the bilayer is akin to the statement that the Hamiltonian commutes with the $x$-Pauli matrix, $\px$:
\begin{align}
\left[ \hat{H}_{\k} ,\hat{\tau}_{x} \right] = 0,\,\,\qquad\text{where}\,\,\, \hat{\tau}_{x} = \begin{pmatrix} 0 & 1 \\ 1&0 \end{pmatrix}
\end{align}
It should be clear that this operation swaps the two planes of the bilayer, and so implements that mirror operation that we are referring to. The addition of a magnetic field $\bm B$ is typically implemented via a Peierls substitution, resulting in a dramatic change to the structure of the Hamiltonian and eigenstates. In particular, working in Landau gauge we only preserve translation invariance in a single direction, so in general the eigenstates will be labeled by a generalized Landau level index, $n$, and transverse momentum, $k_y$. However, as long as the magnetic field does not break this mirror symmetry, i.e. $\bm{B} = B_z \hat{z}$, it remains the case that eigenstates of $\hat H$ are also eigenstates of $\px$, i.e.
\begin{align}
[\hat{H},\px ] &= 0\\
\hat H \lvert n,k_y, \pm \rangle &= E_{(n,k_y,\pm)} \lvert n,k_y, \pm \rangle\\
\px \lvert n,k_y, \pm \rangle &= \pm \lvert n,k_y, \pm \rangle
\end{align}
Note that these are the exact eigenstates of the system, and they are necessarily orthogonal. Also notice that none of these statements depend on the form of the interlayer tunneling $t_{\perp}(\k)$.

The absence of magnetic breakdown is then most easily understood by considering the structure of the energy spectrum. Oscillations in any physical quantity arise because of periodicity in the structure of the energy spectrum as a function of $1/B$. The discrete two-fold mirror symmetry means that the Hamiltonian separates into two independent blocks, so that the energy spectrum for these $+$ and $-$ sectors can be solved independently. Because these sectors can be treated as independent systems, as the magnetic field is varied, each sector produces a \textbf{single} fundamental frequency in quantum oscillations. This results in two (possibly degenerate) quantum oscillation frequencies, with neither magnetic breakdowns nor beat (sum or difference) frequencies.

Fig. ~\ref{fig:symmetricfs} and~\ref{fig:asymmetricfs} provide
confirmation of these symmetry arguments. In
Fig.~\ref{fig:symmetricfs} we have considered identical dispersions
$\varepsilon(\k) = -2t(\cos{k_x} + \cos{k_y}) - \mu$ with $t=1$ and
$\mu = -2.8t$, and $t_{\perp}(\k) = -0.1t \cos^{2}{(2k_y)}$. This
form of the interlayer tunneling is both technically simple to
implement, and produces nodes in the bilayer splitting. As is clear
from the Fourier transform, no magnetic breakdown is present, and
only two fundamental frequencies are seen when the interlayer
tunneling is present. In Fig.~\ref{fig:asymmetricfs} we weakly break
the symmetry by considering dispersions of the form $\varepsilon(\k)
= -2(t_a\cos{k_x} + t_b\cos{k_y}) - \mu$ in one layer, and
$\varepsilon(\k) = -2(t_b\cos{k_x} + t_a\cos{k_y}) - \mu$ in the
next layer, with $t_b = 0.95t_a$. In the absence of interlayer
tunneling, only one frequency is seen in QOs (these pockets have
identical areas), but a finite interlayer tunneling leads to
multiple breakdown orbits.

\section{Recursive Green's function method for the DOS of a tri-diagonal block Hamiltonian}\label{appendix:technical}
As is shown in the main text, the Hamiltonian in $k_y$ and $\sigma$
basis only involves finite-range coupling and is block tri-diagonal
\begin{align}
\hat{H}_{k_y,\sigma}&= \begin{pmatrix}
\ddots &\vdots&&& \\
\ldots & \hat{h}_{x-1,\sigma}& \hat{t} &&\\
& \hat{t}&\hat{h}_{x,\sigma} & \hat{t}&\\
&&\hat{t}&\hat{h}_{x+1,\sigma}&\ldots\\
&&&\vdots&\ddots
\end{pmatrix},\\
\quad \hat{h}_{x,\sigma}&=\begin{pmatrix} 2t_{y,1}\cos(2\pi \Phi x -
k_y) + \frac{4\pi\tilde{g}
B\sigma}{\cos\theta} & t_{\perp}\exp(i 2\pi \Phi a_c x\tan{\theta} )\\
t_{\perp}\exp(-i 2\pi \Phi a_c x\tan{\theta} ) & 2t_{y,2}\cos(2\pi
\Phi x  - k_y)+ \frac{4\pi\tilde{g} B\sigma}{\cos\theta}
\end{pmatrix},\\
\quad \hat{t} &=\begin{pmatrix}
t_{x,1}&0\\
0&t_{x,2}
\end{pmatrix}
\end{align}

We are interested in the DOS $\rho_\sigma(\mu)$ of spin $\sigma$
electrons at chemical potential $\mu$ defined as
\begin{align}
\rho_\sigma(\mu) =  -\frac{1}{\pi L_x
}\text{Tr}\left(\text{Im}[\hat{G}_\sigma(\mu)  ]\right)
\end{align}
\begin{align}
\hat{G}_\sigma(\mu) = \left[ (\mu + i\delta)\mathbb{I} -
\hat{H}_{k_y,\sigma}\right]^{-1}
\end{align}
where we have used the fact that the physical quantities are
independent of $k_y$ in the thermodynamic limit to suppress the
summation over the $k_y$ index.

To obtain the diagonal elements of the Green's function
$\hat{G}_\sigma(\mu)$, we note the inverse of the following block
tri-diagonal matrix may be calculated recursively
\begin{align}
\hat{G}^{-1}_\sigma(\mu) = (\mu + i\delta)\mathbb{I} -
\hat{H}_{k_y,\sigma}=
\begin{pmatrix}
\bm{a}_{1,1}&\bm{a}_{1,2} &&&\\
\bm{a}_{2,1}&\bm{a}_{2,2} & \bm{a}_{2,3} &&\\
&\bm{a}_{3,2}&\bm{a}_{3,3}&\bm{a}_{3,4}&\\
&&\ddots&\ddots&\ddots\\
\end{pmatrix}
\end{align}
where $\bm{a}_{i,i}= (\mu + i\delta)\mathbb{I} - \hat{h}_{x,\sigma}$
and $\bm{a}_{i,i+1}=\bm{a}_{i,i+1}=\hat{t}$. This is accomplished by
the following recursive algorithm, which consists of two independent
sweeps (and hence the computation is linear in the size $L_x$):

For increasing $i =1,2,\ldots,N-1$ we define
\begin{align}
\bm{c}^{L}_i = - \bm{a}_{i+1,i}(\bm{d}^L_{i})^{-1},
\end{align}
with $ \bm{d}^{L}_{1} = \bm{a}_{1,1}$ and $\bm{d}^{L}_{i} =
\bm{a}_{i,i} + \bm{c}^{L}_{i-1}\bm{a}_{i-1,i}$; for decreasing $i
=N,N-1,\ldots,2$ we define
\begin{align}
\bm{c}^{R}_i = - \bm{a}_{i-1,i}(\bm{d}^R_{i})^{-1},
\end{align}
where $\bm{d}^{R}_{N} = \bm{a}_{N,N}$ and $\bm{d}^{R}_{i} =
\bm{a}_{i,i} + \bm{c}^{R}_{i+1}\bm{a}_{i+1,i}$, then the diagonal
blocks of $\hat{G}_\sigma(\mu) = \left[ (\mu + i\delta)\mathbb{I} -
\hat{H}_{k_y,\sigma}\right]^{-1}$ are given by
\begin{align}
\hat G_{i,i} = (-\bm{a}_{i,i} + \bm{d}^{L}_{i} +
\bm{d}^{R}_{i})^{-1},\quad i=1,2,3,\ldots,N
\end{align}

\section{Effective masses of electron pockets and Zeeman splitting coefficient $\tilde{g}$}\label{appendix:meff}

\subsection{Value of $\tilde{g}$ coefficient for Zeeman splitting in our
tight-binding model}

The effective mass of a band structure is defined as
\begin{align}
m^{*} = \frac{\hbar^2}{2\pi} \frac{\partial {S}_{k}}{\partial \mu}
\label{eq:appmeff}
\end{align}
where ${S}_{k}$ is the $k$-space area enclosed by the Fermi surface
at chemical potential $\mu$.

The dispersion relation in our tight-binding model in one of the
single layers is (equivalent to the 3rd orbit in Fig.
\ref{fig:rawFTs})
\begin{equation}
\epsilon_{k}=-2t_{a}\cos k_{x}a-2t_{b}\cos
k_{y}b\eqsim-2t_{a}-2t_{b}+t_{a}k_{x}^{2}a^{2}+t_{b}k_{y}^{2}b^{2}
\label{eq:appdispersion}
\end{equation}
near the bottom of the band, where $a$ and $b$ are the sizes of the
unit cell. At chemical potential $\mu$ the Fermi surface is close to
an ellipsis with $k_{x}^{0}=\sqrt{\frac{\mu}{t_{a}a^{2}}}$ and
$k_{y}^{0}=\sqrt{\frac{\mu}{t_{b}b^{2}}}$, thus the area enclosed by
the Fermi surface
\begin{equation}
S_{k}=\pi k_{x}^{0}k_{y}^{0}=\frac{\pi \mu}{ab\sqrt{t_{a}t_{b}}}
\end{equation}
The effective mass of the model near the band bottom is
\begin{eqnarray}
m^{*}&=&\frac{\hbar^{2}}{2ab\sqrt{t_{a}t_{b}}}
\end{eqnarray}

By definition, the Zeeman splitting is
\begin{equation}
E_{Zeeman}=\pm\frac{g}{2}\mu_{B}B=\pm\frac{g\pi\hbar^{2}}{2abm_{e}}\frac{\Phi}{\cos\theta}=\pm\pi\sqrt{t_{a}t_{b}}\frac{gm^{*}}{m_{e}}\frac{\Phi}{\cos\theta}
\end{equation}
where $\mu_{B}=e\hbar/2m_{e}$ is the Bohr magneton and $\Phi$ is the
dimensionless quantity of the number of magnetic flux quantum
$\Phi_0=h/e$ per $x-y$ plaquette.

Note that $g=2$ for electron spin and $m^{*}/m_{e}\eqsim1.6$ in
YBCO, $t_{a}=1$ and $t_{b}=1/3$,
\begin{equation}
E_{Zeeman}\eqsim\pm 0.92\times 2\pi\Phi/\cos\theta
\end{equation}

In fact, the quadratic approximation in $\epsilon_{k}$ in Eq.
\ref{eq:appdispersion} underestimates the effective mass $m^{*}$ due
to the higher order terms we have neglected. A more careful
treatment and comparison between the numerical and theoretical
$\theta$ dependence suggests the best choice is
\begin{equation}
E_{Zeeman}\eqsim \pm 0.87\times2\pi\Phi/\cos\theta
\end{equation}
suggesting $\tilde{g} = 0.87$ in connection with Eq. \ref{eq:mainham0}.

\subsection{Effective mass for different semiclassical orbits}
\begin{figure}
\begin{center}
\includegraphics[width=0.35\textwidth]{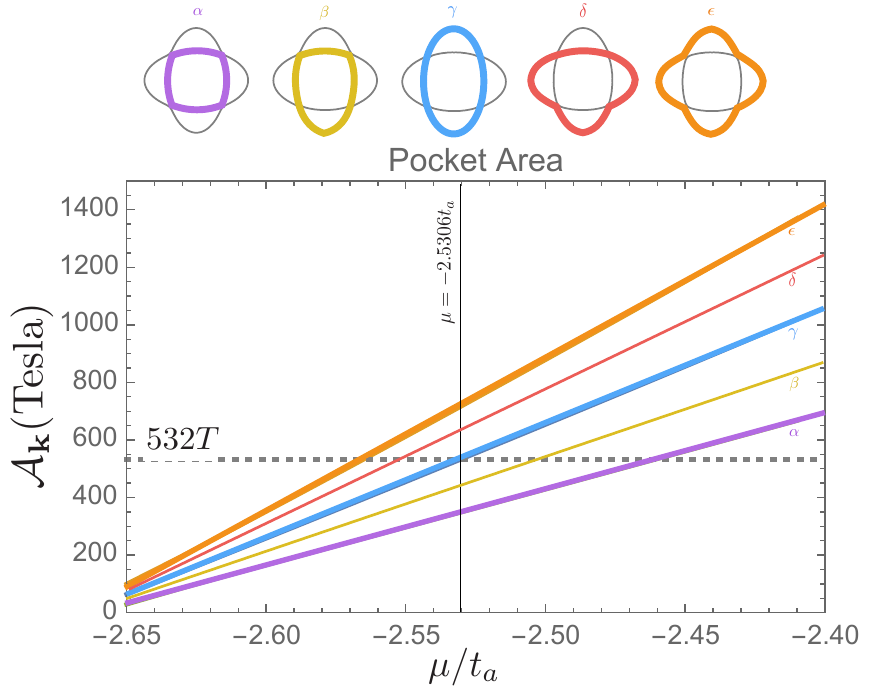}
\caption{Different slopes of $S_{k}$ versus $\mu$ suggest the
effective masses are different for the different orbits. The areas
of $\alpha$, $\gamma$ and $\epsilon$ orbits are obtained from exact
calculations of the Fermi surface, while for $\beta$ and $\delta$
orbits the areas are based on interpolation between the $\alpha$,
$\gamma$ and $\epsilon$ orbits (shown as the thinner lines). The
vertical line is the value of $\mu=-2.5306 t_a$ chosen throughout
our calculations.} \label{fig:meff}
\end{center}
\end{figure}
\FloatBarrier
While $\tilde{g} = 0.87$ determines the effective mass of the electron
pocket in a single layer and the central peak in the QO power
spectrum, it is conceivable that the effective mass of the other
viable semiclassical cyclotron orbits associated with the side peaks
be different, as their enclosed areas are necessarily modified -
Fig.~\ref{fig:meff} shows the enclosed areas of these orbits as the
chemical potential is varied, and the effective mass extracted from
the corresponding slope according to Eq. \ref{eq:appmeff} is fully
consistent with that obtained from the fit to QO amplitude versus
$\theta$ angle of the magnetic field $\vec B$ in Fig.
\ref{fig:waterfall}.

\section{Fourier Transform Analysis}\label{appendix:fourier}

Fast-Fourier transforms of finite data sets are known to introduce
frequency `artifacts' into power-spectrum plots. These artifacts
originate in the choice of how the data is truncated. For example, a
`boxcar' function---whereby the signal is simply truncated at the
start and end---introduces high-frequency components due to the
sharp cutoffs at the data boundaries. Modern signal processing
solves this through `apodization', whereby the data is brought to
zero in some way at the boundary. The choice of apodization function
depends on what features in the data are of interest.
\\

The data in Fig. \ref{fullrange} were processed using a Kaiser
window, designed to resolve closely-spaced frequencies while
suppressing side-lobes (at the expense of absolute amplitude
determination, which was not important for this analysis). The
weighting function $w$ for $N$ data points is defined as
\begin{equation}
w\left(n\right) = \frac{I_0 \left(\pi \alpha \sqrt{1-\left(\frac{2n}{N-1} -1 \right)^2}\right)}{I_0\left(\pi \alpha\right)},
\label{eq:kaiser}
\end{equation}
where $I_0$ is the zeroth modified Bessel function of the first kind
and $\alpha$ controls the roll-off of the weighting function (chosen
to be 1.7 for this work). Fig. \ref{fig:four1} shows the effect of
such a windowing function on a signal and its Fourier transform.
\\

\begin{figure}
\centering
\includegraphics[width=0.48\textwidth]{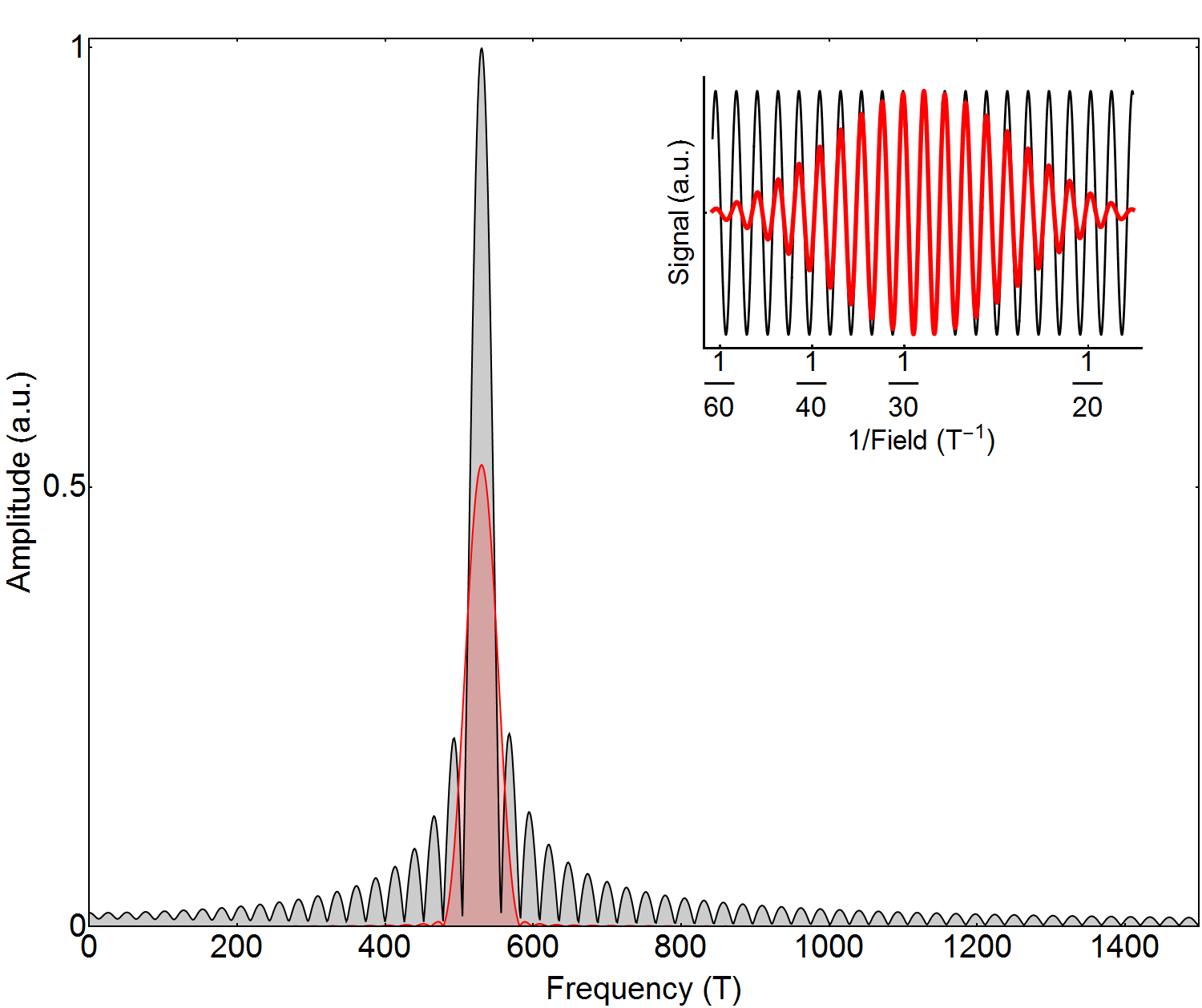}
\caption{The inset shows two simulated data sets: one is apodized with a boxcar function (black), and the other uses equation \ref{eq:kaiser} with $\alpha = 1.7$ (red). The Fourier transform of the boxcar-apodized data shows multiple side-lobes introduced from the sharp cut-off. The data apodized with the Kaiser window has a main peak suppressed by about a factor of two, but with the first side-lobe suppressed 20 times more than that in the box-car data. }
\label{fig:four1}
\end{figure}

Simulated QO data is shown in the inset of Fig. \ref{fig:four2}. The
data contains only the three central frequencies: 440, 530, and 620
T. Specifically, the function is
\begin{equation}
\tau = e^{-150/B}\left(\cos\left(\frac{2 \pi ~440}{B}-\pi\right) + \cos\left(\frac{2 \pi ~530}{B}-\pi\right) +\cos\left(\frac{2 \pi ~620}{B}-\pi\right)  \right).
\label{eq:sim1}
\end{equation}
Note the lack of side-lobes near 350 and 710 T: this demonstrates
that the $\alpha$ and $\epsilon$ peaks in Fig. \ref{fullrange} are
not artifacts of the data analysis.

\begin{figure}
\centering
\includegraphics[width=0.48\textwidth]{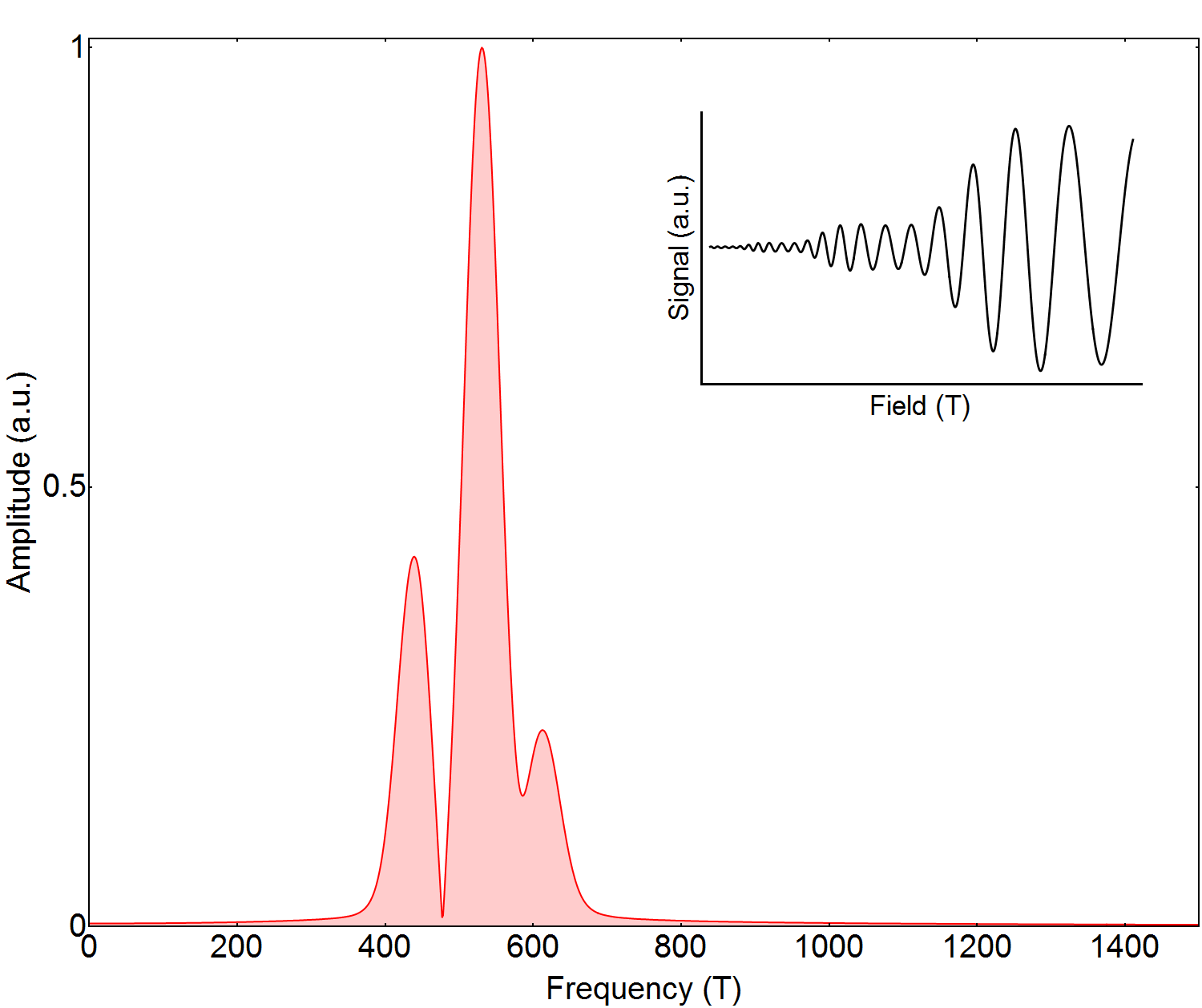}
\caption{The inset shows simulated data from equation \ref{eq:sim1}
before the window is applied. The Fourier transform uses the Kaiser
window with $\alpha = 1.7$---the same as the red curve in Fig.
\ref{fig:four1} and in Fig. \ref{fullrange} in the main text. Note
that there are no extraneous side-lobes.  } \label{fig:four2}
\end{figure}

\end{widetext}

\end{document}